# U.C.L.A. Law Review

## Private Accountability in the Age of Artificial Intelligence

Sonia K. Katyal

**ABSTRACT**

In this Article, I explore the impending conflict between the protection of civil rights and artificial intelligence (AI). While both areas of law have amassed rich and well-developed areas of scholarly work and doctrinal support, a growing body of scholars are interrogating the intersection between them. This Article argues that the issues surrounding algorithmic accountability demonstrate a deeper, more structural tension within a new generation of disputes regarding law and technology. As I argue, the true promise of AI does not lie in the information we reveal to one another, but rather in the questions it raises about the interaction of technology, property, and civil rights.

For this reason, I argue that we are looking in the wrong place if we look only to the state to address issues of algorithmic accountability. Instead, given the state's reluctance to address the issue, we must turn to other ways to ensure more transparency and accountability that stem from private industry, rather than public regulation. The issue of algorithmic bias represents a crucial new world of civil rights concerns, one that is distinct in nature from the ones that preceded it. Since we are in a world where the activities of private corporations, rather than the state, are raising concerns about privacy, due process, and discrimination, we must focus on the role of private corporations in addressing the issue. Towards this end, I discuss a variety of tools to help eliminate the opacity of AI, including codes of conduct, impact statements, and whistleblower protection, which I argue carries the potential to encourage greater endogeneity in civil rights enforcement. Ultimately, by examining the relationship between private industry and civil rights, we can perhaps develop a new generation of forms of accountability in the process.

**AUTHOR**

Haas Distinguished Chair, University of California at Berkeley, and Co-director, Berkeley Center for Law and Technology. Many, many thanks to the following for conversation and suggestions: Bob Berring, Ash Bhagwat, Andrew Bradt, Ryan Calo, Robert Cooter, Jim Dempsey, Deven Desai, Bill Dodge, Dan Farber, Malcolm Feeley, Andrea Freeman, Sue Glueck, Gautam Hans, Woodrow Hartzog, Chris Hoofnagle, Molly van Houweling, Margaret Hu, Gordon Hull, Sang Jo Jong, Neal Katyal, Craig Konnoth, Prasad Krishnamurthy, Joshua Kroll, Amanda Levendowski, David Levine, Mark MacCarthy, Peter Menell, Cal Morrill, Deirdre Mulligan, Tejas Narechania, Claudia Polsky, Julia Powles, Joel Reidenberg, Mark Roark, Bertrall Ross, Simone Ross, Andrea Roth, Pam Samuelson, Jason Schultz, Paul Schwartz, Andrew Selbst, Jonathan Simon, Katherine Strandburg, Joseph Turow, Rebecca Wexler, Felix Wu, and John Yoo. I am tremendously grateful to Danielle Citron, whose commentary at the 2017 Privacy Law Scholars Conference was incredibly helpful. Andrea Hall, Renata Barreto and Aniket Kesari provided extraordinary research assistance.

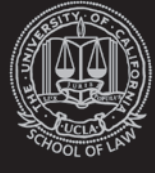

## TABLE OF CONTENTS

## Introduction

Algorithms in society are both innocuous and ubiquitous. They seamlessly permeate both our on– and offline lives, quietly distilling the volumes of data each of us now creates. Today, algorithms determine the optimal way to produce and ship goods, the prices we pay for those goods, the money we can borrow, the people who teach our children, and the books and articles we read—reducing each activity to an actuarial risk or score. "If every algorithm suddenly stopped working," Pedro Domingos hypothesized, "it would be the end of the world as we know it."[1]

Big data and algorithms seem to fulfill modern life's promise of ease, efficiency, and optimization. Yet our dependence on artificial intelligence (AI) does not come without significant social welfare concerns. Recently, a spate of literature from both law reviews and popular culture has focused on the intersection of AI and civil rights, raising traditional antidiscrimination, privacy, and due process concerns.[2] For example, a 2016 report revealed that Facebook used algorithms to determine users' "ethnic affinity," which could only be understood as a euphemism for race.[3] The categories then allowed advertisers to exclude users with certain ethnic affinities from seeing their ads.[4] After initially defending the categories as positive tools to allow users to see more relevant ads, Facebook removed the categories for housing, credit, and employment ads three months later, ostensibly due to antidiscrimination concerns.[5] Despite this move, in September of 2018, the ACLU filed a charge against Facebook with the EEOC, contending that another of its tools violated both labor and civil rights laws by enabling employers to target only men to apply for a wide variety of jobs, including roofing, driving and other opportunities in their advertising.[6] The plaintiffs who came forward included both women and gender nonbinary job seekers who used Facebook in order to receive job ads and other recruitment opportunities, but as their ACLU lawyer explained, they were often hard to identify as plaintiffs. "You don't know, as a Facebook user, what you're not seeing," she explained.[7]

---

1.    Pedro Domingos, The Master Algorithm: How the Quest for the Ultimate Learning Machine Will Remake Our World 1 (2015).

2.    *See* sources cited *infra* notes 18 and 27.

3.    Julia Angwin & Terry Parris, Jr., *Facebook Lets Advertisers Exclude Users by Race*, ProPublica (Oct. 28, 2016, 1:00 PM), http://www.propublica.org/article/facebook-lets-advertisers-exclude-users-by-race [https://perma.cc/ZLU3-N9R8].

4.    *Id.*; *see also* David Lumb, *Facebook Enables Advertisers to Exclude Users by 'Ethnic Affinity'*, Engadget (Oct. 28, 2016), http://www.engadget.com/2016/10/28/facebook-enables-advertisers-to-exclude-users-by-ethnic-affinit [https://perma.cc/2UCD-45LV].

5.    *See Improving Enforcement and Promoting Diversity: Updates to Ads Policies and Tools*, Facebook: Newsroom (Feb. 8, 2017), http://newsroom.fb.com/news/2017/02/improving-enforcement-and-promoting-diversity-updates-to-ads-policies-and-tools [https://perma.cc/3DRQ-4MAM].

Even aside from the allegations of facilitating employment discrimination, that same year, Facebook was rocked by the allegations from a whistleblower, Christopher Wylie, previously at a firm called Cambridge Analytica, who claimed to have come up with the idea to harvest millions of Facebook profiles (50 million approximately), and then target users with political ads that would mesh with their psychological profile.[8] By the 2016 presidential election, Wylie's intervention took on a more sinister cast. Working with an academic, Aleksandr Kogan, millions of people were targeted with fake ads and content, allegedly paid for by Russian organizations.[9] Wylie claimed to have "broke[n] Facebook," and, as the Guardian points out, it was "on behalf of his new boss, Steve Bannon."[10]

Since algorithms tend to show users content that can affirm their existing interests and beliefs,[11] within these filter bubbles, fake news flourished,[12] perhaps affecting the results of the 2016 U.S. election.[13] By coming forward, and telling

---

6.    Nitasha Tiku, *ACLU Says Facebook Ads Let Employers Favor Men Over Women*, WIRED (Sept. 18, 2018, 9:00 AM), https://www.wired.com/story/aclu-says-facebook-ads-let-employers-favor-men-over-women [https://perma.cc/W826-XWFD].

7.    *Id.*

8.    Carole Cadwalladr, *'I Made Steve Bannon's Psychological Warfare Tool': Meet the Data War Whistleblower*, GUARDIAN (Mar. 18, 2018, 5:44 PM), https://www.theguardian.com/news/2018/mar/17/data-war-whistleblower-christopher-wylie-faceook-nix-bannon-trump [https://perma.cc/HK5S-VS5C].

9.    *See* David Folkenflik, *Facebook Scrutinized Over Its Role in 2016's Presidential Election*, NPR (Sept. 26, 2017, 4:59 AM), https://www.npr.org/2017/09/26/553661942/facebook-scrutinized-over-its-role-in-2016s-presidential-election [https://perma.cc/JKA9-XMKF].

10.   Cadwalladr, *supra* note 8.

11.   *See generally* ELI PARISER, FILTER BUBBLE: HOW THE NEW PERSONALIZED WEB IS CHANGING WHAT WE READ AND HOW WE THINK (2012) (describing this phenomenon).

12.   *See* Craig Silverman, *This Analysis Shows How Viral Fake Election News Stories Outperformed Real News on Facebook*, BUZZFEED (Nov. 16, 2016, 5:15 PM), http://www.buzzfeed.com/craigsilverman/viral-fake-election-news-outperformed-real-news-on-facebook [https://perma.cc/Z65E-6ZA8] (showing that Facebook users liked, shared, or commented on the top-performing fake news stories significantly more than the top stories from legitimate news sites).

13.   Today, attempts to address fake news and false information have led to efforts to provide the public with reports of disputed information, such as Facebook's Disputed Flags, small red badges next to potentially untrustworthy sources. In response to criticism that these measures were not sufficient, Facebook has replaced Disputed Flags with Relevant Articles—links that redirect users to high quality, reputable content. *See* Catherine Shu, *Facebook Will Ditch Disputed Flags on Fake News and Display Links to Trustworthy Articles* his story, Wylie became one of the first—and few—tech whistleblowers to risk liability for violating his nondisclosure agreements, triggering a slate of federal inquiries as a result.[14]

Today, for the most part, these reports are the tip of the iceberg regarding the potential impact of algorithmic bias on today's society.[15] But there also is a deeper parallel between civil rights and artificial intelligence that is worth noting. Typically, we think about algorithms in the same way we think about law—as a set of abstract principles manifesting rational objectives. "Math isn't human, and so the use of math can't be immoral," the traditional argument goes.[16]

Yet we now face the uncomfortable realization that the reality could not be further from the truth. The suggestion that algorithmic models are free from social bias represents what has been called an "appeal to abstraction," overlooking concerns that implicate fairness, accountability, and social welfare.[17] These presumptions also overlook the most basic of human costs as well. The idea that algorithmic decisionmaking, like laws, are objective and neutral obscures a complex situation. It refuses to grapple with the causes and effects of systematic and structural inequality, and thus risks missing how AI can have disparate impacts on particular groups. In our zeal to predict who will be the most productive and loyal employee or who will likely execute a terror attack, we collect data on everything. We collect data before we can even conceive of, let alone prove, its relevance—like reading tea leaves before the water has even boiled. We try to predict and preempt things long before they occur, but it can

---

lead to the misapprehension of characteristics, and even worse, a misapplication of stereotypical assumptions.

At first glance, because data collection has now become ubiquitous, the benefits of algorithmic decisionmaking often seem to outweigh their costs. And this is mostly right. Without it those troves of data would remain useless and inscrutable. Yet for members of certain groups, particularly the less wealthy, an algorithm's mistake can be ruinous—leading to denials of employment, housing, credit, insurance, and education.[18] These outcomes demonstrate a central problem in algorithmic accountability: While algorithmic decisionmaking may initially seem more reliable because it appears free from the irrational biases of human judgment and prejudice, algorithmic models are also the product of their fallible creators, who may miss evidence of systemic bias or structural discrimination in data or may simply make mistakes.[19] These errors of omission— innocent by nature—risk reifying past prejudices, thereby reproducing an image of an infinitely unjust world.

Years ago, constitutional law had a similar moment of reckoning. Critical race scholars and others demonstrated how the notion of colorblindness actually obscured great structural inequalities among identity-based categories.[20] The ideals enshrined in our Constitution, scholars argued, that were meant to offer formal equality for everyone were not really equal at all. Rather, far from ensuring equality for all, the notionally objective application of law actually had the opposite effect of perpetuating discrimination toward different groups.

There is, today, a curious parallel in the intersection between law and technology. An algorithm can instantly lead to massive discrimination between groups. At the same time, the law can fail spectacularly to address this discrimination because of the rhetoric of objectivity and secrecy surrounding it. Because many algorithms are proprietary, they are resistant to discovery and scrutiny. And this is one of the central obstacles to greater accountability and transparency in today's age of big data.

This Article argues that the issues surrounding algorithmic accountability demonstrate a deeper, more structural tension within a new generation of disputes regarding law and technology, and the contrast between public and private accountability.  As I argue, the true potential of AI does not lie in the information we reveal to one another, but rather, in the questions they raise about the interaction of technology, intellectual property, and civil rights. Previous literature focused on the relationship between law and technology—which came first, and why.[21]  Commentators lamented the pervasive mismatch between the infinite promise of technology and the comparably more limited reach of law and regulation.[22]  In summing up the view that technology would create a world in which laws would impede with pedestrian concerns, Lawrence Lessig wrote, "[o]verregulation stifles creativity.  It smothers innovation.  It gives dinosaurs a veto over the future.  It wastes the extraordinary opportunity for a democratic creativity that digital technology enables."[23]  Technologists have, and often rightfully so, framed legal regulation—particularly in the world of intellectual property—as outdated, outmoded, and unnecessarily impeding innovation.[24]  Law—particularly intellectual property law—seemed inappropriate and unbearably rigid in its incrementalism and failure to appreciate the possibilities of a digital economy.

---

18. For a detailed discussion of the role of algorithms in society, see CATHY O'NEIL, WEAPONS OF MATH DESTRUCTION: HOW BIG DATA INCREASES INEQUALITY AND THREATENS DEMOCRACY (2016), and FRANK PASQUALE, THE BLACK BOX SOCIETY: THE SECRET ALGORITHMS THAT CONTROL MONEY AND INFORMATION (2015).

19. *See* Oscar H. Gandy, Jr., *Engaging Rational Discrimination: Exploring Reasons for Placing Regulatory Constraints on Decision Support Systems*, 12 ETHICS & INFO. TECH. 29, 30 (2010) (arguing that human-generated data produce biases in automated systems).

20. *See generally* CRITICAL RACE THEORY: THE KEY WRITINGS THAT FORMED THE MOVEMENT (Kimberlé Crenshaw et al. eds., 1995).

Today, we see something quite different.  In the context of artificial intelligence, we see a world where, at times, intellectual property principles prevent civil rights from adequately addressing the challenges of technology, thus stagnating a new generation of civil rights altogether.[25]  Courts all too often defer to AI decisionmaking and deny defendants access to the source code for software that produces the evidence used to convict them.[26]  This new era raises grave civil rights concerns, and yet the law has been woefully inadequate at ensuring greater transparency and accountability.[27]

As I argue, we also need to ask a fundamental question, in each of the contexts we face: Do we need to redesign the algorithm? Or, do we instead need to redesign civil rights law to address the algorithm? Either approach requires very different types of solutions, some of which can be legislated, and some of which cannot. That is precisely why it is so important for us to think broadly and creatively about what the law can and cannot do. We must remember, after all, that far too many acts of AI injustice occur at the hands of private industry, further amplifying the issue of opacity. At the same time, it is also necessary for us not to think of AI as an abstract set of black boxes, but rather as a specific set of granular opportunities for analysis and reflection that can draw on areas of psychology, regulation, and behavioral economics in order to encourage greater transparency.

For this reason, I argue that we are looking in the wrong place if we look to the state alone to address issues of algorithmic accountability. Instead, we must turn elsewhere to ensure more transparency and accountability that stem from private industry, rather than public regulation. That is, of course, not to say that greater regulation requiring transparency is not desirable. However, given the current reluctance of both state and federal legislators to address the challenges posed by AI, it makes sense to explore opportunities for greater endogeneity in addressing civil rights concerns, particularly given the information asymmetry between the industries that design AI and the larger public.

To that end, I divide this Article into four Parts, half descriptive, half normative. Part I explores how machine learning models can unwittingly create skewed results due to well-documented forms of bias in the data that machine learning algorithms are trained upon. Part II turns to the aftermath of algorithmic decisionmaking, drawing on examples from advertising, employment, and price discrimination to show the emergence of civil rights concerns in each context. Finally, in Parts III and IV, I turn to the normative question of how to address the nexus between private corporations and algorithmic accountability. As I argue, the issue of algorithmic bias represents a crucial new world of civil rights concerns, one that is distinct in nature from the ones that preceded it. Since we are in a world where the activities of private corporations, rather than the state, are raising concerns about privacy, due process, and discrimination, we must focus on the role of private corporations in addressing the issue. Here, in the absence of pending government action, I present two potential models to ensure greater transparency, drawn from self-regulation and whistleblower protection, that demonstrate the possibility of greater endogeneity in civil rights enforcement.

---

89 WASH. L. REV. 1, 1 (2014) (discussing the lack of transparency in automated government decisions); Roger Allan Ford & W. Nicholson Price II, *Privacy and Accountability in Black-Box Medicine*, 23 MICH. TELECOMM. & TECH. L. REV. 1 (2016) (healthcare); Brandon L. Garrett, *Big Data and Due Process*, 99 CORNELL L. REV. ONLINE 207 (2014) (positing that there are overlooked issues at the intersection of big data and due process, namely the need for rules around e-discovery and the reconfiguration of *Brady v. Maryland* in the context of government data); Margaret Hu, *Big Data Blacklisting*, 67 FLA. L. REV. 1735 (2015) (administrative proceedings); Jennifer L. Mnookin, *Of Black Boxes, Instruments, and Experts: Testing the Validity of Forensic Science*, 5 EPISTEME 343, 343 (2008) (asserting that courts have accepted superficial explanations behind scientific methods rather than requiring empirical testing and evaluation); Erin Murphy, *The New Forensics: Criminal Justice, False Certainty, and the Second Generation of Scientific Evidence*, 95 CALIF. L. REV. 721, 747–48 (2007) (forensic techniques); Patrick Toomey & Brett Max Kaufman, *The Notice Paradox: Secret Surveillance, Criminal Defendants, & the Right to Notice*, 54 SANTA CLARA L. REV. 843 (2015) (prosecutorial surveillance); Jennifer N. Mellon, Note, *Manufacturing Convictions: Why Defendants Are Entitled to the Data Underlying Forensic DNA Kits*, 51 DUKE L.J. 1097 (2001) (DNA testing protocols).

## I.    DATA AND ITS DISCONTENTS

The Oxford English Dictionary defines an algorithm as "a procedure or set of rules used in calculation and problem-solving."[28] The term originally meant nothing more than basic arithmetic. Now, with the advent of more advanced computers and the ability to collect, compute, and compare ever-larger amounts of data, algorithms represent the promise and peril of social engineering on a scale vaster, yet more precise, than ever possible before. That development is attributable in no small part to the advent of artificial intelligence, which comprises machines that receive inputs from the environment, then learn from or interpret those inputs, and then potentially take certain actions or decisions that affect the environment.[29] Although the machine creates an illusion of autonomy, its actions depend completely on the code that humans write for it.

---

28.    *Algorithm*, OXFORD ENGLISH DICTIONARY (3d ed. 2012), http://www.oed.com/view/Entry/4959?redirectedFrom=algorithms (last visited Oct. 13, 2018).

29.    This definition is drawn from a definition offered by Stuart Russell and Peter Norvig. *See* Daniel Faggella, *What Is Artificial Intelligence? An Informed Definition*, TECHEMERGENCE (May 15, 2017), http://www.techemergence.com/what-is-artificial-intelligence-an-informed-definition [https://perma.cc/L6KQ-WRED]. The term artificial intelligence (AI) was coined by John McCarthy for the Dartmouth Conferences in 1956. He defined it as "the science and engineering of making intelligent machines." John McCarthy, *Basic Questions*, STAN. COMPUTER SCI DEP'T: FORMAL REASONING GROUP (Nov. 12, 2007), http://www-formal.stanford.edu/jmc/whatisai/node1.html [https://perma.cc/X8VA-VDC2]. Today, some scholars observe that the term AI comprises two different branches of entities—"smart" computers (like deep learning), and an unrealized "artificial general intelligence," (or AGI). *Id.*

Algorithms result from a complex interaction of features, classifications, and targets, all of which draw upon a maze of hazy interactive and embedded values.[30] According to Tarleton Gillespie, "the algorithm comes after the generation of a 'model,' i.e. the formalization of the problem and the goal in computational terms."[31] So, for example, consider the goal of "giving a user the most relevant search results for their queries.[32] That would require, Gillespie explains, a model that efficiently calculated "the combined values of pre-weighted objects in the index database, in order to improve the percentage likelihood that the user clicks on one of the first five results."[33] The resulting algorithm would comprise a series of steps that aggregated the values in an efficient manner, and something that might deliver rapid results.[34] What makes something algorithmic, he explains, is that the result is produced by an information system "that is committed (functionally and ideologically) to the computational generation of knowledge or decisions."[35]

The use of mathematical principles to solve social problems is nothing new. Parole boards have used actuarial models to assess the risk of recidivism with varying degrees of sophistication since the 1920s.[36] Advanced computing and its ability to collect, compute, and compare ever-larger amounts of data have also allowed algorithms to grow more complex and powerful.[37] Where Stewart Brand could once curate and compile the Whole Earth, Google now promises its algorithm will do the same—but better.[38] Its search algorithm, Google claims, is mere math; thus, its sorting and filtering is impartial and produces the most

---

relevant, useful results.[39] Those more relevant results, in turn, attract more users, which allow Google to sell its ad space at a premium.[40] Similarly, since its inception in a Seattle garage, Amazon has used algorithms to quantify consumer preferences and thus recommend and sell products, often to its comparative advantage.[41] Netflix, too, uses an algorithm to compare a viewer's habits to those of others.[42] And OkCupid once dominated online dating with its algorithms before everyone switched to Tinder, allowing users to simply "swipe right" over another individual's picture as a way of indicating interest in that individual.[43] Target famously used algorithms to create predictive models so accurate, it could tell a teenager was pregnant before her family knew.[44]

While the effects of algorithms' predictions can be troubling in themselves, they become even more problematic when the government uses them to distribute resources or mete out punishment.[45] The Social Security Administration uses algorithms to aid its agents in evaluating benefits claims; the Internal Revenue Service uses them to select taxpayers for audit; the Food and Drug Administration uses algorithms to study patterns of foodborne illness; the Securities and Exchange Commission uses them to detect trading misconduct;

---

39.  Paško Bilić, *Search Algorithms, Hidden Labour and Information Control*, 2016 BIG DATA & SOC'Y 1, 3 (discussing how Google operates).

40.  Craig E. Wills & Can Tatar, *Understanding What They Do With What They Know*, *in* PROCEEDINGS OF THE 2012 WORKSHOP ON PRIVACY IN THE ELEC. SOC'Y (Oct. 15, 2012).

41.  *See, e.g.*, Andrea M. Hall, Note, *Standing the Test of Time: Likelihood of Confusion in* Multi Time Machine v. Amazon, 31 BERKELEY TECH. L.J. 815, 827–30 (2016); Julia Angwin & Surya Mattu, *Amazon Says It Puts Customers First. But Its Pricing Algorithm Doesn't*, PROPUBLICA (Sept. 20, 2016, 8:00 AM), http://www.propublica.org/article/amazon-says-it-puts-customers-first-but-its-pricing-algorithm-doesnt [https://perma.cc/W6JU-CU2Z]; Franklin Foer, *Amazon Must Be Stopped*, NEW REPUBLIC (Oct. 9, 2014), http://newrepublic.com/article/119769/amazons-monopoly-must-be-broken-radical-plan-tech-giant [https://perma.cc/PVR4-GDEH].

42.  *See* Ashley Rodriguez, *How Netflix (NFLX) Determines What to Pay for Shows and Films*, QUARTZ (Dec. 27, 2016), http://qz.com/872909 [https://perma.cc/HUF6-CYN9].

43.  *See* Chava Gourarie, *Investigating the Algorithms That Govern Our Lives*, COLUM. JOURNALISM REV. (Apr. 14, 2016), http://www.cjr.org/innovations/investigating_algorithms.php [https://perma.cc/7SRT-8ENY] (noting use of algorithms in online dating services); Benjamin Winterhalter, *Don't Fall in Love on OkCupid*, JSTOR DAILY (Feb. 10, 2016), https://daily.jstor.org/dont-fall-in-love-okcupid [https://perma.cc/775E-FAK7].

44.  Solon Barocas & Helen Nissenbaum, *Big Data's End Run Around Anonymity and Consent*, *in* PRIVACY, BIG DATA, AND THE PUBLIC GOOD 44, 62 (Julia Lane et al. eds., 2014); Charles Duhigg, *How Companies Learn Your Secrets*, N.Y. TIMES (Feb. 16, 2012), http://www.nytimes.com/2012/02/19/magazine/shopping-habits.html.

45.  *See* Charlie Beck & Colleen McCue, *Predictive Policing: What Can We Learn From Wal-Mart and Amazon About Fighting Crime in a Recession?*, POLICE CHIEF (Nov. 2009), http://www.policechiefmagazine.com/predictive-policing-what-can-we-learn-from-wal-mart-and-amazon-about-fighting-crime-in-a-recession [https://perma.cc/8Y89-VRGH] (example of advocacy for further government use of algorithms).

local police departments employ their insights to predict the emergence of crime hotspots; courts use them to sentence defendants; and parole boards use them to decide who is least likely to reoffend.[46]

Algorithms hold tremendous value. Big data promises significant benefits to the economy, allowing consumers to find and sort products more quickly, which in turn lowers search costs. Yet their potential to shape society in dramatic, unanticipated ways has often been underestimated. The dominant perception is that algorithms are but simple mathematical principles, rearranged to reveal patterns and make predictions. Who would quibble, the reasoning goes, that one plus one is two? Under this view, objectivity seemingly benefits users. Instead of weighing the credibility of and comparing various answers, algorithms reveal the single best answer. Algorithmic recommendations consequently can save users' search and information costs by tailoring services and content to consumers.[47] AI can also aid in the detection of financial mismanagement, identity theft, and credit card fraud.[48]

Now that we have more data than ever, proponents suggest, the results of predictive analytics should be better, more robust, and more accurate than ever before.[49] Algorithmic decisionmaking, through rote analysis of quantitative information, seemingly creates an appealing alternative to human judgments that risk subjectivity and bias.[50] And, to be fair, most applications of algorithms do in fact seem relatively harmless. Is it really so bad if Facebook's news algorithm shows me the latest in adorable kitten news rather than Syrian refugee updates, most people might ask? Maybe that is what I want, a typical consumer

---

46. Ronald Bailey, *Welcoming Our New Algorithmic Overlords?*, REASON (Oct. 1, 2016), https://reason.com/archives/2016/10/01/welcoming-our-new-algorithmic [https://perma.cc/YV7L-RK8N]; *see also* Jon Kleinberg et al., *Prediction Policy Problems*, 105 AM. ECON. REV.: PAPERS & PROC. 491, 494–95 (2015) (discussing how improved prediction techniques using machine learning can have significant policy implications).

47. *See generally* Elizabeth J. Altman et al., *Innovating Without Information Constraints: Organizations, Communities, and Innovation When Information Costs Approach Zero, in* THE OXFORD HANDBOOK OF CREATIVITY, INNOVATION, AND ENTREPRENEURSHIP 353 (Christina E. Shalley et al. eds., 2015).

48. Anjanette H. Raymond et al., *Building a Better HAL 9000: Algorithms, the Market, and the Need to Prevent the Engraining of Bias*, 15 NW. J. TECH. & INTELL. PROP. 215, 217 (2018).

49. For an example of an optimistic view, see Ric Simmons, *Quantifying Criminal Procedure: How to Unlock the Potential of Big Data in Our Criminal Justice System*, 2016 MICH. ST. L. REV. 947 (2016).

50. *See* Nathan R. Kuncel, Deniz S. Ones & David M. Klieger, *In Hiring, Algorithms Beat Instinct*, HARV. BUS. REV. (May 1, 2014), http://hbr.org/2014/05/in-hiring-algorithms-beat-instinct [https://perma.cc/84BK-F23U].

might reason. It hardly seems worse than what a panel of human news editors could choose.[51]

Indeed, that is how many people encounter algorithms: innocent enhancements of a consumer experience. Algorithms, however, do much more—both by addressing, analyzing, and then potentially replicating our worlds of implicit bias. And the results can often be mystifying. When an algorithm produces a t-shirt that says "keep calm and rape a lot"[52] or Twitter users transform an innocent chatbot into a white supremacist in less than a day,[53] there is clearly more at stake than innocent depictions. Here, machine learning can mirror back to us a particularly uncomfortable construction of reality.[54] To take one example, a recent study has argued that as a machine learning model acquires capabilities that approximate the context of human language—a process known as "word embedding"—it demonstrates, and replicates, the same troubling implicit biases that we see in human psychology.[55] For example, the same study showed that the words "female" and "woman" were more closely associated with the domestic sphere, and occupations associated with the arts and humanities, as opposed to the terms "male" and "man," which were closer to professions associated with mathematics and engineering.[56]

---

51.   *See* RISJ Admin, *Brand and Trust in a Fragmented News Environment*, REUTERS INST., https://reutersinstitute.politics.ox.ac.uk/our-research/brand-and-trust-fragmented-news-environment [https://perma.cc/MG6G-VZDU]; *see also* Steven Porter, *Can Facebook Resolve Its News Problems Without Losing Credibility?*, CHRISTIAN SCI. MONITOR (Jan. 11, 2017), http://www.csmonitor.com/Business/2017/0111/Can-Facebook-resolve-its-news-problems-without-losing-credibility [https://perma.cc/VK3S-A93F] (describing trade-offs between human versus algorithmic editing of Facebook's news feed).

52.   Chris Baraniuk, *The Bad Things That Happen When Algorithms Run Online Shops*, BBC: FUTURE (Aug. 20, 2015), http://www.bbc.com/future/story/20150820-the-bad-things-that-happen-when-algorithms-run-online-shops [https://perma.cc/9XRN-FR4T].

53.   James Vincent, *Twitter Taught Microsoft's AI Chatbot to Be a Racist Asshole in Less Than a Day*, VERGE (Mar. 24, 2016, 6:43 AM), http://www.theverge.com/2016/3/24/11297050/ tay-microsoft-chatbot-racist [https://perma.cc/PS9V-2PP6].

54.   *See* Farhad Manjoo, *How Netflix Is Deepening Our Cultural Echo Chamber*s, N.Y. TIMES (Jan. 11, 2017), http://www.nytimes.com/2017/01/11/technology/how-netflix-is-deepening-our-cultural-echo-chambers.html.

55.   *See* Aylin Caliskan, Joanna J. Bryson & Arvind Narayanan, *Semantics Derived Automatically From Language Corpora Contain Human-like Biases*, 356 SCIENCE 6334 (2017); *see also* Hannah Devlin, *AI Programs Exhibit Racial and Gender Biases, Research Reveals*, GUARDIAN (Apr. 13, 2017, 2:00 PM), http://www.theguardian.com/ technology/2017/apr/13/ai-programs-exhibit-racist-and-sexist-biases-research-reveals [https://perma.cc/4DWS-M6NR].

56.   Devlin, *supra* note 55. As Amanda Levendowski explains, Google's use of word2vec, a word embedding toolkit, reflects "the gendered bias embedded in the Google News corpus used to train it," offering the example of the toolkit projecting "that man is to computer programmer in the same way that woman is to homemaker." Amanda Levendowski, *How Copyright Law Can Fix Artificial Intelligence's Implicit Bias Problem*, 93 WASH. L. REV. 579,

Errors at any stage can become amplified in the next stage, producing deviant outcomes in complex, troubling, and sometimes difficult-to-detect ways. Since algorithmic models reflect the design choices of the humans who built them, they carry the biases of the observer or instrument.[57] The sheer size of big data also obscures smaller variations.[58] While most researchers focus on the dangers of false positives and false negatives in data,[59] far more pernicious types of discrimination can result from how classes of data are defined, and the sorts of examples and rules that algorithms learn from such data.[60] In an excellent study, Solon Barocas and Andrew Selbst detailed a number of different ways in which the data mining process can give rise to models that risk having an adverse impact on protected classes, stemming from biased data inputs, measurement errors or missing variables, or inappropriate uses of criteria that also serve as proxies for a protected class or group.[61] In the Subparts below, I analyze the potential interplay of both statistical and cognitive forms of bias, and discuss how each can affect the design of the algorithm, the data it is trained upon, and ultimately its outcome.[62]

---

581 (2018); *see also* Raymond et al., *supra* note 48, at 218–19 (noting similar concerns regarding race, in addition to gender).

57. *See* James Bogen, *Theory and Observation in Science*, STAN. ENCYCLOPEDIA PHIL. (Jan. 11, 2013), http://plato.stanford.edu/archives/spr2013/entries/science-theory-observation (noting how various philosophers "cast suspicion on the objectivity of observational evidence by challenging the assumption that observers can shield it from biases . . . ."); Tyler Woods, '*Mathwashing,' Facebook and the Zeitgeist of Data Worship*, TECHNICAL.LY BROOKLYN (June 8, 2016, 9:18 AM), http://technical.ly/brooklyn/2016/06/08/fred-benenson-mathwashing-facebook-data-worship [https://perma.cc/BR7W-CG44].

58. *See* Brooke Foucault Welles, *On Minorities and Outliers: The Case for Making Big Data Small*, 1 BIG DATA & SOC'Y 1 (2014), (discussing some problems that arise with big data).

59. *See, e.g.*, *Data Science—Dealing With False Positives and Negatives in Machine Learning*, TERADATA: DATA SCI. BLOG (Dec. 28, 2015, 12:51 PM), http://community.teradata.com/t5/Learn-Data-Science/Data-Science-Dealing-with-False-Positives-and-Negatives-in-ba-p/79675 [https://perma.cc/Q89M-DTRV].

60. *See, e.g.*, Solon Barocas & Andrew D. Selbst, *Big Data's Disparate Impact*, 104 CALIF. L. REV. 671, 680 (2016); *see also* James Grimmelmann & Daniel Westreich, *Incomprehensible Discrimination*, 7 CALIF. L. REV. ONLINE 164 (2016) (exploring issues of accountability and transparency in machine learning); Joshua A. Kroll et al., *Accountable Algorithms*, 165 U. PA. L. REV. 633, 680 (2017) ("These decision rules are machine-made and follow mathematically from input data, but the lessons they embody may be biased or unfair nevertheless.").

61. Barocas & Selbst, *supra* note 60, at 677.

62. Ricardo Baeza-Yates, *Bias on the Web*, 61 COMM. ACM 54, 54 (2018) (defining statistical bias to comprise "[A] systematic deviation caused by an inaccurate estimation or sampling process."); *see also* Barocas & Selbst, *supra* note 60, at 677.

## A.    Statistical and Historical Bias in Big Data

As Kate Crawford and Meredith Whittaker have observed in the inaugural AI Now Report, bias in big data generally results from one of two causes.[63]  The first is largely internal to the process of data collection—when errors in data collection, like inaccurate methodologies, lead to inaccurate depictions of reality.[64]  The second type, however, comes from an external source.  It happens when the underlying subject matter draws on information that reflects or internalizes some forms of structural discrimination and thus biases the data as a result.[65]  Imagine, they explain, a situation where data on job promotions might be used to predict career success, but the data was gathered from an industry that systematically promoted men instead of women.[66]  While the first kind of bias can often be mitigated by "cleaning the data" or improving the methodology, the latter might require interventions that raise complex political ramifications because of the structural nature of the remedy that is required.[67]  As a result of these issues, bias can surface in the context of input bias level (when the source data is biased because it may lack certain types of information), training bias (when bias appears in the categorization of the baseline data), or through programming bias (when bias occurs from a smart algorithm learning and modifying itself from interaction with human users or incorporating new data).[68]

Although mathematical algorithms have been around for thousands of years, today, machine learning algorithms are trained on a body of data that is selected by designers or by past human practices.  This process is the "learning" element in machine learning; the algorithm learns, for example, how to pair

---

queries and results based on a body of data that produced satisfactory pairs in the past.[69] The quality of a machine learning algorithm's results often depends on the comprehensiveness and diversity of the data that it digests.[70] Bad data, in other words, can perpetuate inequalities through machine learning, leading to a feedback loop that replicates existing forms of bias, potentially impacting minorities as a result. For example, recently, the first international beauty contest derived from AI sparked controversy after the results from its 6000 entries (from over one hundred countries) revealed that of the forty-four winners, nearly all were white.[71] Why? Although there are probably a host of reasons, the main problem was that the training data that was used, ostensibly to establish standards of attractiveness among humans, did not include enough people of color.[72]

## 1. Underrepresentation and Exclusion

One common form of machine learning is supervised learning (where one has input variables and output variables, and an algorithm is used to train the machine to learn the mapping function from the input to the output).[73] But there is also unsupervised machine learning, where instead, we rely on a machine to identify patterns in data instead, using insights from statistics and neuroscience.[74]

---

69. *See id.* For more information on types of machine learning, see also Edwards & Veale, *supra* note 37, at 25–27.
70. *See* Barocas & Selbst, *supra* note 60, at 688.
71. *See* Sam Levin, *A Beauty Contest Was Judged by AI and the Robots Didn't Like Dark Skin*, GUARDIAN (Sept. 8, 2016, 6:42 PM), http://www.theguardian.com/technology/2016/sep/08/artificial-intelligence-beauty-contest-doesnt-like-black-people [https://perma.cc/5T3P-F2DW].
72. *Id.* (citing Alex Zhavoronkov, Beauty.AI's chief science officer).
73. For a fuller explanation, see Jason Brownlee, *Supervised and Unsupervised Machine Learning Algorithms*, MACHINE LEARNING MASTERY (Mar. 16, 2016), http://machinelearningmastery.com/supervised-and-unsupervised-machine-learning-algorithms [https://perma.cc/R58R-EWG2].
74. For more discussion of these methods of unsupervised learning, see Martin Hynar, Michal Burda & Jana Šarmanová, *Unsupervised Clustering With Growing Self-Organizing Neural Network—A Comparison With Non-Neural Approach, in* PROCEEDINGS OF THE 2005 DATABASES, TEXTS, SPECIFICATIONS AND OBJECTS (DATESO) WORKSHOP 58 (2005). *See also* Nikki Castle, *Supervised vs. Unsupervised Machine Learning*, DATASCIENCE.COM (July 13, 2017), http://www.datascience.com/blog/supervised-and-unsupervised-machine-learning-algorithms [https://perma.cc/9YPW-8RJY]; Bernard Marr, *Supervised V Unsupervised Machine Learning—What's The Difference?*, FORBES (Mar. 16, 2017, 3:13 AM), http://www.forbes.com/sites/bernardmarr/2017/03/16/supervised-v-unsupervised-machine-learning-whats-the-difference/2 (explaining difference between the two).

With supervised learning, since machine learning is based on a system of patterns and correlations in the data to make predictions, these predictions can often be inaccurate if the training data is unrepresentative of the general population that is being studied.[75]  Moreover, there may be noise in the training data itself, stemming from inaccurate information about individuals in the population.[76]  In addition, choices that are made by humans—what features should be used to construct a particular model, for example—can comprise sources of inaccuracy as well.[77]  An additional source of error can come from the training of the algorithm itself, which requires programmers to decide, essentially, how to weigh sources of potential error.[78]

Further, the quality of data can be affected by practices such as excluding outliers, or editing, cleaning, or mining data.[79]  As Solon Barocas and Andrew Selbst have argued:

> Data mining can go wrong in any number of ways.  It can choose a target variable that correlates to [a] protected class more than others would, reproduce the prejudice exhibited in the training examples, draw adverse lessons about protected classes from an unrepresentative sample, choose too small a feature set, or not dive deep enough into each feature.  Each of these potential errors is marked by two facts: the errors may generate a manifest disparate impact, and they may be the result of entirely innocent choices made by data miners.[80]

Since minority interpretations of a search term, for example, do not help Google show relevant ads, generate clicks, or produce revenue on a mass scale, Google and its counterparts might ignore or minimize them in their search results and queries.[81]  In other words, depending on what a company is looking to market, these outliers are simply deviations from the valuable average and therefore excluded.  Their uncommon traits can become lost and ignored in the sea of big data.

Other types of errors have everything to do with the impact of categorization.  Categorization, while key to an algorithmic model's success, can

---

also be its greatest downfall, because it can miss evidence of structural discrimination and bias. As Tarleton Gillespie has written, "[c]ategorization is a powerful semantic and political intervention: what the categories are, what belongs in a category, and who decides how to implement these categories in practice, are all powerful assertions about how things are and are supposed to be."[82] To demonstrate, Gillespie offers the example of a situation involving Amazon in 2009, when nearly 57,000 gay-friendly books disappeared from its sales lists, because they had been wrongly characterized as "adult" books.[83] The error revealed that Amazon had been programming its machine learning model to calculate "sales rank" by excluding adult books from consideration.[84] While the idea of excluding adult books from sales lists might make intuitive sense (since there may be some restrictions on age-related purchases), the model failed to grapple with a known problem in society, which is that often things that are characterized as "adult" or "obscene" are LGBT-related, when the same behavior in an opposite sex context is not classified in the same manner. As a result, a mistake such as Amazon's can have dramatic effects on the visibility of resources for individual consumers seeking validation and community through the consumption of LGBT-related texts. This categorization not only adds to a problematic invisibility of gay texts, but it also feeds into an invisibility of *consumers* of these texts.

Ricardo Baeza-Yates, in a powerful article, describes how common issues like self-selection bias, activity bias, cultural and cognitive bias can skew research on Web-based activities.[85] Aside from these sorts of bias, data collected on the Web is drawn from a skewed demographic, since it favors those with educational, economic, technological, and even linguistic advantages (since over 50 percent of the most popular websites are in English, when only 13 percent of the world speaks English).[86] Elsewhere, in the context of health and big data, researchers have reported a troubling homogeneity among big data.[87] It turns out, some analysts argue, that big data has failed to include marginalized communities, including African American, Latino, Native American populations, people of a lower socioeconomic status, LGBT individuals, and immigrants.[88] Not only are these people disproportionately missing from sources like internet histories, social media, and credit card usage, but they are also missing from electronic health records and genomic databases.[89]

---

82.  Gillespie, *supra* note 30, at 171.
83.  *Id.*
84.  *Id.*
85.  Baeza-Yates, *supra* note 62, at 56 (citing studies that reflect that on Facebook, a large dataset shows that only 7 percent of users produce 50 percent of the posted content; on Amazon, only 4 percent of active users produce the posted reviews; and on Twitter, only 2 percent produced 50 percent of the posts).
86.  *Id.* at 57.
87.  *See* Sarah E. Malanga et al., *Who's Left Out of Big Data? How Big Data Collection, Analysis, and Use Neglect Populations Most in Need of Medical and Public Health Research and Interventions, in* Big Data, Health Law, and Bioethics 98, 98–99 (I. Glenn Cohen et al. eds., 2018).

Further, even the techniques of data collection can bias results. Easily available data tends to be reported and analyzed more often, leading to a reporting bias because harder to find information may never make it into the dataset.[90] There are classic examples of selection bias, where some individuals are picked for study rather than others. But there is also exclusion bias, resulting when individuals are excluded from certain studies, as discussed above.[91] Results can even differ based on whether something is written or oral (modality bias).[92] Baeza-Yates describes an additional level of bias that can also be unwittingly introduced by interaction designers, who might create bias in designing the user's interface; in one example, he points out that content that is placed in the top left corner of the screen tends to attract more eyes and clicks, a type of "position bias."[93] Ranking bias is a related form of bias, which privileges top-ranked items over ones that are lower in the order of relevance.[94]

And, in turn, the effects of misrepresentation can impact different groups. In other words, if the machine learning model is trained on data that is biased in some way, then decisions that are derived from that data can systematically disadvantage individuals who happen to be over– or underrepresented in the dataset.[95] As Baeza-Yates concludes, "[b]ias begets bias."[96] Here, depending on the issue, data mining can actually resurrect past prejudices if it relies on prior decisions that are already rife with discrimination. In one example, a UK hospital developed a computer program to sort medical school applicants. However, the program relied on its prior decisions, which had systematically been shown to discriminate against women and minority applicants with the same credentials as other applicants, thus risking the same outcome here.[97] Preexisting past biases in datasets, then, can lead to the reconstruction and replication of bias in the future, creating forms of second-order bias as a result.[98]

---

88. *See id.*

89. *Id.*

90. For a definition of reporting bias, see *Reporting Bias: Definition and Examples, Types*, STAT. HOW TO (Oct. 12, 2017), http://www.statisticshowto.com/reporting-bias [http://perma.cc/4QHU-6TLZ]. *See also* Jonathan Gordon & Benjamin Van Durme, *Reporting Bias and Knowledge Acquisition*, 2013 PROCEEDINGS OF THE WORKSHOP ON AUTOMATED KNOWLEDGE BASE CONSTRUCTION 25 (analyzing generally how reporting bias functions in artificial intelligence).

91. *See, e.g.*, Miguel Delgado-Rodríguez & Javier Llorca, *Bias*, 58 J. EPIDEMIOLOGY & COMMUNITY HEALTH 635, 637 (2004) (describing bias in an epidemiological context); Joyce Chou, Oscar Murillo & Roger Ibars, *How to Recognize Exclusion in AI*, MEDIUM (Sept. 26, 2017), https://medium.com/microsoft-design/how-to-recognize-exclusion-in-ai-ec2d6d89f850 [http://perma.cc/L3S3-JZT3] (discussing examples of exclusion bias).

92. *See* Mark L. Elliott et al., *Modality Effects in Short Term Recognition Memory*, 94 AM. J. PSYCHOL. 85 (1981).

93. Baeza-Yates, *supra* note 62, at 58.

94. *Id.*

95. Barocas & Selbst, *supra* note 60, at 680–81.

96. Baeza-Yates, *supra* note 62, at 60.

The problem is not just one of inadequate representation. The model's determination or conclusions may also fail to communicate some recognition of its own risks of inaccuracy, leading to the risk of overconfidence in its results and a failure to communicate attendant ambiguity.[99] As a result, as Joshua Kroll and his coauthors indicate, there are a variety of antidiscrimination implications that arise from choices of inputs. One might use membership in a protected class directly or rely on data that is insufficiently representative of a protected class (for example, relying on employment data that is historically biased against women to assess female applicants). Or it might use factors that may be proxies for protected class membership (for example, length of tenure may seem like a benign category, but women who leave the workplace due to childrearing may lower the average tenure for all women, risking disparate impact if tenure serves as a proxy for gender).[100]

However, these issues are often exceedingly difficult to locate and to address. It is difficult to eliminate proxies if they provide valuable information, and it is often difficult, ex post, to improve the quality of data relied upon.[101] As one commentator, Matthew Carroll explains, "[t]he average engineer is not thinking about bias or proper provenance when designing a neural network."[102] He continues:

> They are focused on nuances such as ideal network topology, activation functions, training gradients, weight normalization, and data overfitting. Once a model is trained, engineers quite frequently lack understanding of the model's actual decision-making process. What if they're called on to explain why a model made the decision that it did—to prove, for example, it didn't make a legally questionable decision, like discriminating on the basis of race? What if a data subject seeks to exercise her right to prevent her data from being used to train the model, or used in the model at all, a right protected by the EU's primary data protection regulation, the GDPR? This is where today's governance models start to break down.[103]

---

97. Barocas & Selbst, *supra* note 60, at 682.
98. Baeza-Yates, *supra* note 62, at 60.
99. *See* Andrea Roth, *Machine Testimony*, 126 YALE L.J. 1972, 1992–93 (2017).
100. Kroll et al., *supra* note 60, at 681.
101. *Id.*
102. Matthew Carroll, *The Complexities of Governing Machine Learning*, DATANAMI (Apr. 27, 2017), http://www.datanami.com/2017/04/27/complexities-governing-machine-learning [https://perma.cc/7GPJ-KUNG].

While researchers might correctly argue that some algorithmic models do not explicitly consider protected identity characteristics in their predictions and are quantitative in nature, they may ignore existing, implicit biases that stem from potential proxies and algorithms' potential to exacerbate them.[104] Jeremy Kun offers another example of how minorities can get treated unfairly by describing something called the "sample size" problem, which researcher Moritz Hardt described as a tendency for statistical models about minorities to perform worse than models that predict behavior of the general population.[105] And if that mathematical minority aligns with a racial minority, then the algorithm might completely ignore an entire racial population. As Kun writes, "an algorithm with 85% accuracy on US participants could err on the entire black sub-population and still seem very good."[106]

## 2. Oversurveillance in Data Selection and Design

If the prior discussion focused on the risks of exclusion from statistical and historical underrepresentation, this Subpart focuses on the opposite risk of overrepresentation, which can also lead to imprecise perceptions and troubling stereotypes. Here, an algorithmic model might associate certain traits with another unrelated trait, triggering extra scrutiny of certain groups.[107] In such cases, it can be hard to prove discriminatory intent in the analysis; just because an algorithm produces a disparate impact on a minority group, it does not always mean that the designer intended this result.[108]

---

103. *Id.*
104. This can be a classic result of anchoring bias—focusing on one aspect of information that fails to take into account other variables, such as structural discrimination. *See* Amos Tversky & Daniel Kahneman, *Judgment Under Uncertainty: Heuristics and Biases*, 185 SCIENCE 1124, 1128–30 (1974).
105. *See* Hardt, *supra* note 81; *see also* Jeremy Kun, *supra* note 16 (discussing Hardt).
106. Kun, *supra* note 16.
107. *See* FED. TRADE COMM'N, BIG DATA: A TOOL FOR INCLUSION OR EXCLUSION? 9 (2016) ("[W]ith large enough data sets, one can generally find some meaningless correlations."); Martin Frické, *Big Data and Its Epistemology*, 66 J. ASS'N INFO. SCI. & TECH. 651, 659 (2015) (discussing the possibility of spotting new patterns in data).
108. *See* Kroll et al., *supra* note 60, at 693–94.

Consider, for example, the debates over predictive policing algorithms. Brett Goldstein, a former officer with the Chicago Police Department and now a public policy scholar, used an algorithm to analyze the locations of prior arrests to predict the location of criminals, a strategy that has been strongly criticized by civil rights groups.[109] Another scholar, Miles Wernick at the Illinois Institute of Technology, developed a program that generated a "heat list" of 400 individuals that had the highest chance of committing a violent crime.[110] He insisted the model was unbiased because it did not rely on any racial, neighborhood, or related information. Instead, he used data about the number and frequency of previous crimes.

Despite his efforts, Wernick's model and its predictions perpetuated existing systemic biases, even where the data analyzed was seemingly unbiased.[111] Individuals who committed prior crimes were detected more often than other potential criminals.[112] Why? Their race or location were more likely to be surveilled.[113] In other words, they were the most likely to get caught because these other characteristics made the individuals more vulnerable to suspicion.[114] The predictive policing algorithm suffered from classic detection bias. Its sample population was far more likely to be surveilled than other social groups, thus overestimating a propensity towards crime.

Frequently, however, the criteria that algorithms consider when making their predictions are secret. Wernick, for instance, refuses to reveal which factors his proprietary algorithm uses, even while touting the accuracy of the police department's list of roughly 400 people most likely to shoot or be shot.[115] As of May 2016, more than 70 percent of the people who had been shot in the city that year were on the list, according to the Chicago police, as were more than 80 percent of those arrested in connection with shootings.[116] However, the same algorithm also led a police commander to turn up at the home of a 22-year-old black man who lived on the south side of Chicago.[117] The police commander warned him not to commit further crimes, although the man had not committed any recent crimes and did not have a violent criminal record.[118]

---

109.  *See* Joshua Brustein, *The Ex-Cop at the Center of Controversy Over Crime Prediction Tech*, BLOOMBERG (July 10, 2017 2:00 AM) https://www.bloomberg.com/news/features/2017-07-10/the-ex-cop-at-the-center-of-controversy-over-crime-prediction-tech [https://perma.cc/9Y6V-QLXL] (discussing Goldstein's predictive policing strategies and related critiques); Kun, *supra* note 16 (same); *see also* Cathy O'Neil, *Gillian Tett Gets It Very Wong on Racial Profiling*, MATHBABE (Aug. 25, 2014), https://mathbabe.org/2014/08/25/gilian-tett-gets-it-very-wrong-on-racial-profiling [https://perma.cc/WRE6-ZHQH] (discussing predictive policing).

110.  Kun, *supra* note 16.

111.  *Id.; see also* Aziz Z. Huq, *Racial Equity in Algorithmic Criminal Justice*, 68 DUKE L.J. (forthcoming 2019); Julia Angwin & Jeff Larson, *Bias in Criminal Risk Scores Is Mathematically Inevitable, Researchers Say*, PROPUBLICA (Dec. 30, 2016, 4:44 PM), http://www.propublica.org/article/bias-in-criminal-risk-scores-is-mathematically-inevitable-researchers-say [https://perma.cc/58UU-3JVY].

112.  Kun, *supra* note 16; *see also* Matt Stroud, *The Minority Report: Chicago's New Police Computer Predicts Crimes, But Is It Racist?*, VERGE (Feb. 19, 2014, 9:31 AM), http://www.theverge.com/2014/2/19/5419854/the-minority-report-this-computer-predicts-crime-but-is-it-racist [https://perma.cc/5AEV-MFSH] (discussing Wernick's work).

113.  *See* Jessica Saunders et al., *Predictions Put Into Practice: A Quasi-Experimental Evaluation of Chicago's Predictive Policing Pilot*, 12 J. EXPERIMENTAL CRIMINOLOGY 347, 356–67 (2016).

114.  *Id.*

In response to questions about such false positives, the Chicago police will say only that the program considers whether an individual's criminal "trend line" is increasing, whether he has been shot before, and whether he has ever been arrested on weapons charges, to make its predictions.[119] They will not reveal the model, nor allow anyone on the list to challenge the factors or the data that it considers.[120] It is easy to see, however, how such questions might readily become a proxy for race, gender, or geography. Residents of neighborhoods that have more crime are more likely to be shot. These neighborhoods consequently become more policed. Because the areas are more policed, police are more likely to detect weapons offenses there and arrest their residents on weapons charges.[121]

---

115.   *See id.* at 15; Saunders et al., *supra* note 113, at 366; Stroud, *supra* note 112 (noting Wernick's reluctance to share specific details about the algorithm).

116.   Monica Davey, *Chicago Policy Try to Predict Who May Shoot or Be Shot*, N.Y. Times (May 23, 2016), http://www.nytimes.com/2016/05/24/us/armed-with-data-chicago-police-try-to-predict-who-may-shoot-or-be-shot.html.

117.   Lum & Isaac, *infra* note 123, at 15.

118.   *Id.*; *see also* Jeremy Gorner, *Chicago Police Use 'Heat List' as Strategy to Prevent Violence*, Chi. Trib. (Aug. 21, 2013), http://www.chicagotribune.com/news/ct-xpm-2013-08-21-ct-met-heat-list-20130821-story.html [https://perma.cc/9PXM-AG3E].

119.   *See* Davey, *supra* note 116.

120.   Even the former White House suggests that transparency about data is essential to effective community policing despite obvious problems with how crime data is collected. *See* Press Release*,* The White House, Office of the Press Sec'y, Fact Sheet: White House Police Data Initiative Highlights New Commitments (Apr. 21, 2016), http://www.whitehouse.gov/the-press-office/2016/04/22/fact-sheet-white-house-police-data-initiative-highlights-new-commitments [http://perma.cc/827N-5VY3]; *see generally* Michael D. Maltz, U.S. Dep't of Justice, Bureau of Justice Statistics, Bridging Gaps in Police Crime Data (1999), http://www.bjs.gov/content/pub/pdf/bgpcd.pdf [https://perma.cc/73ZK-NWL2].

121.   A recent study of Chicago's algorithmic model confirmed that this trend was likely the case, concluding that being on the list correlated only with being arrested for a shooting, not with being the victim of a shooting, as the department had claimed. Saunders et al., *supra* note 113, at 363–64.

Despite these risks, many police departments use software programs to predict crime.[122] The PredPol algorithm uses only three data points—past type, place, and time of crime—to identify the times and places where future crimes are most likely to occur.[123] Critics point out that PredPol and similar algorithms predict not so much where future crime is most likely to occur, but where police are most likely to detect future crime.[124] In other words, PredPol predicts not so much crime as policing. In this respect, algorithmic policing becomes a self-fulfilling prophecy in poor and minority neighborhoods: More policing leads to more arrests, more scrutiny, and potentially greater penalties. The surge in arrests spurs the algorithm to predict a greater need for policing in the same area, leading two scholars to conclude that this was a perfect example of "selection bias meets confirmation bias."[125] Because the algorithms learn from previous arrest data that might reflect biases, they created a feedback loop that perpetuates those biases, even despite their claims to exclude race, gender, or geography from its data.[126]

And this tendency can carry grave results. As Bernard Harcourt has argued, predictive policing can lead to a misdirection of resources, leading crime to be suppressed in areas where individuals are targeted (and thereby receive more resources), at the cost of increasing crime in areas that receive less surveillance (and therefore less resources).[127]

## B.    Errors of Attribution, Prediction, and Preference

Aside from issues with collecting and refining data, cognitive and other forms of implicit bias can also seriously impact both algorithmic design and the data that the algorithm is trained upon. As Christine Jolls, Cass Sunstein, and Richard Thaler argued some time ago, individuals display bounded rationality, bounded willpower, and bounded self-interest—each of which present trajectories that diverge from conventional economic models.[128]    These behavioral trajectories, they argued, require the development of new predictive models to take these biases into account in order to increase their precision.[129]

---

122.  *See generally* G. O. Mohler et al., *Randomized Controlled Field Trials of Predictive Policing*, 110 J. AM. STAT. ASS'N 1399 (2015).

123.  Kristian Lum & William Isaac, *To Predict and Serve?*, SIGNIFICANCE, Oct. 2016, at 14, 17–18, http://rdcu.be/1Ug9.

124.  *See id.* at 18.

125.  *See id.* at 16–19; *see also* Tal Z. Zarsky, *Transparent Predictions*, 2013 U. ILL. L. REV. 1503, 1510 (discussing role of prediction in big data).

126.  *See* Lum & Isaac, *supra* note 123, at 15–16.

127.  *See* BERNARD E. HARCOURT, AGAINST PREDICTION: PROFILING, POLICING, AND PUNISHING IN AN ACTUARIAL AGE 111–38 (2007) (discussed in Simmons, *supra* note 49, at 957). Whether or not crime overall actually decreases depends on the comparative elasticity of each group. Simmons, *supra* note 49, at 957.

The same might be true here, where our cognitive biases might require a much more rigorous interrogation of the ways in which AI can replicate these biases.[130]    Our reliance on heuristics—mental shortcuts to ease and speed decisions—can contribute to the opacity of the problem.[131]    Thus, much more work on implicit bias is necessary to understand how it can be linked to machine learning, data quality, and algorithmic design.    In this Subpart, I explore three specific types of biases—attributional errors, predictive errors, and preference-related errors—to show how they can contribute to the problem of biasing both the design of an algorithm and the data AI relies upon.

## 1.    Attributional Errors

Because we strive to conserve analytic power, we assume that a single example represents a whole class, or we accept the first thought that comes to mind, failing to make adjustments later, because our minds remain anchored to that initial thought.[132]    Aside from shortcuts, we might be directed to process information differently in the presence of emotions, noise, motivations, or other complex factors like social influence in our decisionmaking processes.[133]    In such cases, both the designer of the algorithm—and the subjects represented within the data—can reflect forms of implicit bias that are difficult to detect.

For a moment, think of all of the information about ourselves we gladly hand over to computers.    Monitors, like smartwatches and phones, track our height, weight, where we are, where we are going, how quickly, where, and how much we sleep.    Search engines likewise know all of our questions and their answers, for better or for worse.[134]    But our own cognitive biases already warp what we believe merits recording, what questions are worth asking, and what answers are desirable. Although cognitive bias can assume many forms, relying on self-selected or self-reported data can easily replicate biases on a large scale, due to the simple human cognitive errors that statistics and probability aim to avoid.[135] And when algorithms train on imperfect data, or are designed by individuals who may be unconsciously biased in some manner, the results often reflect these biases, often to the detriment of certain groups.[136] Kate Crawford has described this as AI's "White Guy Problem," referring to the way in which bias becomes reified in the form of AI that draws upon biased data.[137] "Like all technologies before it," she writes, "artificial intelligence will reflect the values of its creators. So inclusivity matters—from who designs it to who sits on the company boards and which ethical perspectives are included. Otherwise, we risk constructing machine intelligence that mirrors a narrow and privileged vision of society, with its old, familiar biases and stereotypes."[138]

---

128. *See* Christine Jolls, Cass R. Sunstein & Richard Thaler, *A Behavioral Approach to Law and Economics*, 50 STAN. L. REV. 1471, 1477–79 (1998).

129. *See id.* at 1477.

130. *Cf.* JUDGMENT UNDER UNCERTAINTY: HEURISTICS AND BIASES (Daniel Kahneman, Paul Slovic & Amos Tversky eds., 1982) (discussing role of heuristics and the bias they produce in human decisionmaking).

131. Martin Hilbert, *Toward a Synthesis of Cognitive Biases: How Noisy Information Processing Can Bias Human Decision Making*, 138 PSYCHOL. BULL. 211, 212–13 (2012).

132. *Id.* at 213.

133. *Id.*

134. *See* Nate Anderson, *Why Google Keeps Your Data Forever, Tracks You With Ads*, ARS TECHNICA (Mar. 8, 2010, 6:20 AM), http://arstechnica.com/tech-policy/2010/03/google-keeps-your-data-to-learn-from-good-guys-fight-off-bad-guys [https://perma.cc/4FPF-FXKK].

Yet studying these biases is key to understanding how to correct or how to qualify our results. Consider, for example, the fact that many individuals make attributional errors, which can affect explanations for a particular phenomenon. Confirmation biases, for example, often problematically lead people to cherry pick data that appears to support their beliefs.[139] Our judgments and answers might also differ depending on how we frame or present a question (framing effect).[140] Likewise, our belief in our control over a dimension can also bias our assessment of that dimension (illusion of control bias).[141] We might inadequately assess our future selves and our needs, thoughts, and preferences

---

(projection bias).[142]  Sometimes we overestimate things like our social desirability; other times, we underestimate it.

And even more problematically, our ego often leads us to be overly confident in our judgments, which can make us loath to reconsider and recalibrate them at a later date.[143]  For example, quantitative models often restrict self-reporting to a limited number of variables, thus simplifying the complexity of a person's lived experience to a set schema.[144]  As a result, researchers might overlook alternate causes of a phenomena because of the variables they have excluded.  They might label a feature that only correlates with another feature as a defining factor in causing the latter, leading to a classic host of errors associated with attribution.[145]  Restricting answers to conform with an observer's expectation, formalized in an algorithm, results in data further confirming those expectations.[146]

Stereotyping is a classic example of this problem.[147]  Princeton Review, for instance, seemed to rely on stereotypes about Asians when it charged zip codes with large Asian populations higher prices for its test preparation.[148]  Although Facebook preferred to describe its racial classifications as "Ethnic Affinity,"[149] it demonstrates the risk of racial or ethnic stereotyping in data aggregation because it allowed marketing executives to choose whether to include or exclude ethnic groups from seeing particular advertising.[150]

Other kinds of biases stem from more subtle forms of stereotyping.  Researchers have documented that individuals treat those who belong to their own social or ethnic group better than those who do not, so-called ingroup

---

bias.[151] Relatedly, we also recognize variation among members of our own group with greater subtlety than members of other groups, referred to as outgroup bias.[152] Although we often think of ourselves as unpredictable and capable of change, we might characterize others as much more predictable, or the reverse (trait ascription bias.).[153]

## 2.    Predictive and Preference-Related Errors

Aside from attributional errors, individuals make qualitative and quantitative predictive errors, leading individuals to mistake correlation for causation. Sometimes we overestimate the probability of events happening based on what happened most recently,[154] view things in the past differently (hindsight bias),[155] or we may construct preferences based on the present moment rather than over time (current moment bias).[156] Other times, we rely heavily on our own beliefs, and then those beliefs gain more and more traction over time—especially if they are adopted by more and more people—leading to a bandwagon effect that obscures the actual events or the cause.[157]

An additional cluster of cognitive bias involves preference-related errors that can often involve incorrect or illogical estimations of value or quality.[158] For example, the data surrounding fake news suggests individuals' focus on familiar information that confirms their existing beliefs. Whether we already agree with or oppose a fact or viewpoint often involves whether it conforms with our expectations (selective perception).[159] Similarly, there is also the illusory truth effect, which leads us to often think things are true because we have heard them before, not because of any actual validity.[160]

---

151.  *Cf.* Michael J. Bernstein et al., *The Cross-Category Effect: Mere Social Categorization Is Sufficient to Elicit an Own-Group Bias in Face Recognition*, 18 PSYCHOL. SCI. 706 (2007) (discussing role of ingroup bias).

152.  For a discussion of ingroup and outgroup bias, see S. Alexander Haslam, Penelope J. Oakes, John C. Turner & Craig McGarty, *Social Identity, Self-Categorization, and the Perceived Homogeneity of Ingroups and Outgroups: The Interaction Between Social Motivation and Cognition*, *in* 3 HANDBOOK OF MOTIVATION & COGNITION: THE INTERPERSONAL CONTEXT 182 (Richard M. Sorrentino & E. Tory Higgins eds., 1996); Donald M. Taylor & Janet R. Doria, *Self-Serving and Group-Serving Bias in Attribution*, 113 J. SOC. PSYCHOL. 201 (1981).

153.  *See* Daniele Kammer, *Differences in Trait Ascriptions to Self and Friend: Unconfounding Intensity From Variability*, 51 PSYCHOL. REP. 99 (1982).

154.  *See* Amos Tversky & Daniel Kahneman, *Availability: A Heuristic for Judging Frequency and Probability*, 5 COGNITIVE PSYCHOL. 207 (1973). For more information on "recency bias" and its applicability to current issues, see Carl Richards, *Tomorrow's Market Probably Won't Look Anything Like Today*, N.Y TIMES (Feb. 13, 2012, 12:23 PM), https://bucks.blogs.nytimes.com/2012/02/13/tomorrows-market-probably-wont-look-anything-like-today.

155.  *E.g.*, Neal J. Roese & Kathleen D. Vohs, *Hindsight Bias*, 7 PERSP. ON PSYCHOL. SCI. 411 (2012).

156.  *See* Shane Frederick et al., *Time Discounting and Time Preference: A Critical Review*, 40 J. ECON. LITERATURE 351, 352 (2002) (referencing "the preference for immediate utility over delayed utility").

157.  *See* Richard Nadeau et al., *New Evidence About the Existence of a Bandwagon Effect in the Opinion Formation Process*, 14 INT'L POL. SCI. REV. 203 (1993).

158.  For a discussion of the role of preference-related error, see Robert E. Scott, *Error and Rationality in Individual Decisionmaking: An Essay on the Relationship Between Cognitive Illusions and the Management of Choices*, 59 S. CAL. L. REV. 329 (1986).

At this point, it may seem as though the grab bag of biases just described may never find their way into an algorithmic function. Yet the truth is that many of them do replicate themselves in data, particularly self-reported data. As Andrea Roth has explained in the context of criminal law, just as humans exhibit hearsay dangers of insincerity, loss of memory, and misperception, the risk of machines replicating these errors can lead to the prospect of a falsehood by design, leading to the misanalysis of an event in court.[161] These dangers can be further magnified if we depend so excessively on AI (automation bias) that we will not be able to detect or correct error.[162] And this can also affect the design of the algorithm itself, including whether or not it can be redesigned to take these issues into account. For example, if studies demonstrate that resumes from European American-sounding names are 50 percent more likely to receive an interview invitation than if the name is African American-sounding in nature, then an algorithm will demonstrate the same social predispositions unless specifically programmed to recognize this disparity.[163]

## C.    Prediction and Punishment: An Example

The dangers that flow from biased data are perhaps best illustrated by the prodigious work of other scholars exploring its role in the criminal justice system. Long before the tech industry fell in love with AI, criminal law scholars had been drawing on behavioral science to explore the costs and benefits of risk

---

prediction.[164]  Today, machine learning and big data play a powerful role in policing strategies.[165]  Rebecca Wexler describes a host of other technologies in the criminal law context that are already being used for forensic purposes and are considered to be proprietary—algorithms that generate candidates for latent fingerprint analysis; to search ballistic information databases for firearm and cartridge matches; and facial recognition technologies, to take just a few examples.[166]

Here, algorithms that are used to sentence defendants or parole prisoners raise issues of racial bias.[167]  For example, in work discussing the Post Conviction

---

164. *See* Malcolm M. Feeley & Jonathan Simon, *The New Penology: Notes on the Emerging Strategy of Corrections and Its Implications*, 30 CRIMINOLOGY 449, 452 (1992) (discussing implications of actuarial assessment in the criminal justice system).

165. *See generally* Andrew Guthrie Ferguson, *Predictive Policing and Reasonable Suspicion*, 62 EMORY L.J. 259 (2012); Erin Murphy, *Databases, Doctrine, and Constitutional Criminal Procedure*, 37 FORDHAM URB. L.J. 803 (2010); Rich, *supra* note 27 (all describing use of these strategies and their implications).

166. Wexler, *supra* note 26, at 1347, 1363–64.

167. *See* Erica Meiners, *How "Risk Assessment" Tools Are Condemning People to Indefinite Imprisonment*, TRUTHOUT (Oct. 6, 2016), http://www.truth-out.org/news/item/37895-how-risk-assessment-tools-are-condemning-people-to-indefinite-imprisonment [https://perma.cc/Y6WA-8LNE].  Algorithms have pervaded the criminal justice system. Sonja Starr's excellent work has demonstrated how evidence-based sentencing (EBS) has raised substantial constitutional concerns.  Sonja B. Starr, *Evidence-Based Sentencing and the Scientific Rationalization of Discrimination*, 66 STAN. L. REV. 803 (2014) [hereinafter Starr, *Evidence-Based Sentencing*]; Sonja B. Starr, *The New Profiling: Why Punishing Based on Poverty and Identity Is Unconstitutional and Wrong*, 27 FED. SENT'G REP. 229 (2015).  For a good discussion of the Starr Stanford paper and its implications, see Luis Daniel, *The Dangers of Evidence-Based Sentencing*, NYU: GOVLAB (Oct. 31, 2014), http://thegovlab.org/the-dangers-of-evidence-based-sentencing [https://perma.cc/KM5L-MG2V].  Others have raised similar concerns.  *See, e.g.*, Melissa Hamilton, *Risk-Needs Assessment: Constitutional and Ethical Challenges*, 52 AM. CRIM. L. REV. 231 (2015) (expressing constitutional concerns); R. Karl Hanson & David Thornton, *Improving Risk Assessments for Sex Offenders: A Comparison of Three Actuarial Scales*, 24 LAW & HUM. BEHAV. 119 (2000); Bernard E. Harcourt, *Risk as a Proxy for Race: The Dangers of Risk Assessment*, 27 FED. SENT'G REP. 237 (2015); Ian Kerr, *Prediction, Pre-emption, Presumption: The Path of Law After the Computational Turn*, *in* PRIVACY, DUE PROCESS AND THE COMPUTATIONAL TURN 91 (Mireille Hildebrandt & Katja de Vries eds., 2013); John Monahan & Jennifer L. Skeem, *Risk Redux: The Resurgence of Risk Assessment in Criminal Sanctioning*, 26 FED. SENT'G REP. 158 (2014); Jonathan Remy Nash, *The Supreme Court and the Regulation of Risk in Criminal Law Enforcement*, 92 B.U. L. REV. 171 (2012); J.C. Oleson, *Risk in Sentencing: Constitutionally Suspect Variables and Evidence-Based Sentencing*, 64 S.M.U. L. REV. 1329 (2011); Dawinder S. Sidhu, *Moneyball Sentencing*, 56 B.C. L. REV. 671 (2015); Roger K. Warren, *Evidence-Based Sentencing: The Application of Principles of Evidence-Based Practice to State Sentencing Practice and Policy*, 43 U.S.F. L. REV. 585 (2009); Danielle Citron, *(Un)Fairness of Risk Scores in Criminal Sentencing*, FORBES (July 13, 2016, 3:26 PM), http://www.forbes.com/sites/daniellecitron/2016/07/13/ unfairness-of-risk-scores-in-criminal-sentencing [https://perma.cc/2PQN-TVHN].  Here, artificial intelligence's "White Guy Problem" becomes reified in the form of algorithms, that risk

Risk Assessment (PCRA) instrument, several scholars have shown some potential for disparate impact based on race,[168] gender,[169] and age.[170]  A recent ProPublica report studied Correctional Offender Management Profiling for Alternative Sanctions (COMPAS), one of the most popular algorithms that is used to assess a defendant's risk of recidivism and subsequently sentence that defendant based on this risk.[171]   Although the creator of the algorithm, Northpointe, developed COMPAS in the late 1990s[172]  to assess risk factors in correctional populations and to provide decision support for case planning and management, rather than sentencing, it has now become a quite powerful tool in assessing four different kinds of risk: general recidivism, violent recidivism, noncompliance, and failure to appear.[173]

---

Northpointe has revealed that the COMPAS analysis considers a subject's basic demographic information,[174] criminal record, and whether anyone in the subject's family has ever been arrested among its 137 questions.[175] Northpointe will not disclose: (1) how the analysis for each of these types of risk varies, (2) all of the factors COMPAS considers, and (3) how it weighs those factors against each other.[176] Some of the questions ask if the subject's parents are divorced, if their parents are incarcerated, what the subject's high school grades were, if they got into a lot of fights in high school, if they have friends who use drugs, and if they have a phone at home.[177] The questions also include moral hypotheticals, like whether the subject agrees or disagrees that "[a] hungry person has a right to steal."[178] It also invites the person administering the questionnaire to speculate on whether or not the subject presents as a gang member.[179]

Although these questions do not necessarily in themselves reveal a bias—because Northpointe refuses to reveal how the algorithm weighs these answers—the only way to assess the algorithm's bias is through its results.[180] ProPublica studied the sentencing of 7000 defendants in Florida's Broward County, obtaining their risk scores and comparing the predictions to how many were charged with new crimes over the next two years (the same benchmark relied upon by the algorithm).[181] When ProPublica tested the proprietary algorithm used to predict recidivism, it discovered that the scores were wrong almost 40 percent of the time, and gravely biased against black defendants, who were "falsely labeled future criminals at almost twice the rate of white defendants."[182] Out of every five people Northpointe predicted would commit another violent crime, the study found that only one actually did.[183] Notably, "[t]he formula was particularly likely to falsely flag black defendants as future criminals, wrongly labeling them this way at almost twice the rate as white defendants."[184]

---

174. *See Algorithms in the Criminal Justice System*, ELECTRONIC PRIVACY INFO. CTR. (Nov. 11, 2017) https://epic.org/algorithmic-transparency/crim-justice/ [https://perma.cc/UUG7-HDT2] ("COMPAS, created by the for-profit company Northpointe, assesses variables under five main areas: criminal involvement, relationships/lifestyles, personality/attitudes, family, and social exclusion."). Starr has pointed out that Northpointe has devised a separate set of question for women. She discusses the constitutional implications of this differential usage by the state in Sonja B. Starr, *Evidence-Based Sentencing*, *supra* note 167, at 823–29, 823 n.76.

175. *See* NORTHPOINTE, RISK ASSESSMENT, http://assets.documentcloud.org/documents/2702103/Sample-Risk-Assessment-COMPAS-CORE.pdf [https://perma.cc/U6XM-DQWY].

176. *See Algorithms in the Criminal Justice System*, *supra* note 174 ("Northpointe has not shared how its calculations are made but has stated that the basis of its future crime formula includes factors such as education levels and whether a defendant has a job.").

177. *Id.*

178. *Id.*

179. Jason Tashea, *Risk-Assessment Algorithms Challenged in Bail, Sentencing, and Parole Decisions*, ABA J. (Mar. 1 2017), http://www.abajournal.com/magazine/article/algorithm_bail_sentencing_parole [https://perma.cc/L53D-63FN]; *see also* NORTHPOINTE, PRACTITIONER'S GUIDE TO COMPAS CORE § 5.1 (2015), http://epic.org/algorithmic-transparency/crim-justice/EPIC-16-06-23-WI-FOIA-201600805-COMPASPractitionerGuide.pdf [https://perma.cc/VP22-L8XA].

180. Northpointe insists, "[t]here's no secret sauce to what we do; it's just not clearly understood . . . ." Tashea, *supra* note 179.

181. Angwin et al., *supra* note 171, at 1 (explaining methodology and results); *see also* Angwin & Larson, *supra* note 111 (discussing implications of findings).

Scores that algorithms like COMPAS produce should comprise only one part of a judge's or parole board's decision. COMPAS, for example, was created not for its applicability in sentencing decisions, but actually to assist probation officers in selecting particular types of treatment in probation decisions.[185] However, it is hard not to imagine that these scores will play an overly significant role in sentencing and ultimately produce harsher sentences for black defendants.

Despite the problems that the ProPublica study documented, the Wisconsin Supreme Court upheld the use of COMPAS in sentencing in July 2016.[186] In 2013, Eric Loomis was charged with crimes related to a drive-by shooting in La Crosse, Wisconsin.[187] He pleaded no contest to a vehicle charge and guilty to eluding an officer.[188] The court ordered a presentencing investigation, including a COMPAS risk assessment report that labeled Loomis a high risk for pretrial recidivism risk, general recidivism risk, and violent recidivism risk.[189] The judge sentenced him to eleven years, explicitly citing the

high score that COMPAS had assigned to him.[190]  Loomis appealed the sentence on due process grounds.[191]

The Wisconsin Supreme Court found that the sentence and the circuit court's reliance on COMPAS did not violate Loomis's rights because he knew the factors that COMPAS considered.  The court pointed out that "Northpointe's 2015 Practitioner's Guide to COMPAS explains that the risk scores are based largely on static information (criminal history), with limited use of some dynamic variables (i.e. criminal associates, substance abuse)."[192]  "[T]o the extent that Loomis's risk assessment is based upon his answers to questions and publicly available data about his criminal history," the court found that Loomis could verify the accuracy of his answers.[193]  The court, however, never addressed that Loomis could not examine the extent those answers had in his risk score because Northpointe guards that information as a trade secret.[194]  Northpointe's guide might offer a tidy explanation for its algorithm and the psychological and sociological theories underpinning it.  There is, however, no way to check that COMPAS actually carries out those theories in a statistically sound and logical manner.

Other courts have also dealt with similar questions.  In *Malenchik v. State*, the Indiana Supreme Court upheld the use of risk assessment scores, reasoning that the scores were a statistically valid means to forecast recidivism, and could be used to supplement a judge's evaluation when used in conjunction with other sentencing evidence to determine an individualized calculation.[195]  In contrast, the Court of Appeals of Indiana in *Rhodes v. State* had expressed concern that the use of a standardized scoring model undercut the trial court's responsibility to craft an individualized sentence.[196]

Recently, a bill introduced in Congress aimed to mandate the procedural use of algorithms to assess the risk of recidivism in parole decisions at federal prisons.[197] As proponents of their predictive utility suggest, these characteristics do not exist in a vacuum.[198] Other criminal protocols, like the extreme vetting protocols and other methods of database screening developed in the wake of Trump's Executive Orders on the Muslim Ban, have led one scholar, Margaret Hu, to refer to them as a system of "Algorithmic Jim Crow."[199]

---

190.  *Id.* at 755, 756 n.18.
191.  *Id.* at 757; *see also* Mitch Smith, *In Wisconsin, a Backlash Against Using Data to Foretell Defendants' Futures*, N.Y. TIMES (June 22, 2016), https://www.nytimes.com/2016/06/23/us/backlash-in-wisconsin-against-using-data-to-foretell-defendants-futures.html.  Loomis argued the sentencing decision violated his right to due process because: (1) Northpointe would not reveal the source code so its validity could not be tested, (2) the judge relied on COMPAS's generalized risk based on defendants like Loomis, rather than considering him as an individual, and (3) the tool improperly considers gender in determining risk.  *Loomis*, 881 N.W.2d at 757.
192.  *Loomis*, 881 N.W.2d at 761.
193.  *Id.*
194.  *See id.* ("[Northpointe] does not disclose how the risk scores are determined or how the factors are weighed.").
195.  928 N.E.2d 564, 575 (Ind. 2010).
196.  896 N.E.2d 1193, 1195 (Ind. Ct. App. 2008), *disapproved of by Malenchik*, 928 N.E.2d at 573.

## II.    THE AFTERLIFE OF THE ALGORITHM

The previous Part outlined some of the ways in which data can be flawed and lead to skewed results due to forms of statistical and cognitive bias. These results can disadvantage a variety of different populations, some of whom might fall within legally protected categories, while others who do not. While the previous set of concerns stemmed from the inputs that algorithms rely upon and some of their limitations, this Part focuses instead on the real-life effects that AI can produce, drawing upon examples from private employment, advertising, and price discrimination.[200]

In a now seminal account, Batya Friedman and Helen Nissenbaum described three central types of bias in computer systems.[201] The first type, also discussed in the previous Part, involved what they described as "preexisting bias," which can reflect the personal biases of individuals who play a significant role in designing the system, either the client or the system designer.[202] This type of bias, they explain, can either be explicit or implicit, and can enter into a system, even despite the best of intentions.[203] A second type of bias, they explain, stems from technical bias, which could include limitations in the hardware, software, or peripherals; "the process of ascribing social meaning to algorithms developed

---

out of context"; or, as they eloquently describe, "when we quantify the qualitative, discretize the continuous, or formalize the nonformal."[204] But a third type of bias, what they call "emergent bias," is harder to detect because it appears only after the design has been completed.[205] Consider an example:

> Using the example of an automated airline reservation system, envision a hypothetical system designed for a group of airlines all of whom serve national routes. Consider what might occur if that system was extended to include international airlines. A flight-ranking algorithm that favors on-line flights when applied in the original context with national airlines leads to no systematic unfairness. However, in the new context with international airlines, the automated system would place these airlines at a disadvantage and, thus, comprise a case of emergent bias.[206]

While Friedman and Nissenbaum may not have noted it at the time, their notion of emergent bias almost perfectly captures the risks inherent in machine learning, where preexisting biases can merge with preexisting and technical biases, producing dynamic results that can disadvantage particular groups.

This risk is particularly pronounced when we consider the degree to which decisions in private industry rule our everyday lives. For example, algorithmic researchers have reported that the ability to draw fine-grained distinctions between individuals through collective risk management strategies can lead to adverse selection among populations and individuals within pooled areas of resources, like insurance.[207] In the case of health insurance, these distinctions may lead to higher premiums through price discrimination strategies.[208] While these strategies have long been in existence, the added reliance on machine learning exacerbates the risk of incomplete or inaccurate judgments and predictions based on the fallibility of the data it relies upon.

At least one author, Cathy O'Neil, has argued that algorithms have a particular disparate impact on the poor because wealthier individuals are more likely to benefit from personal input.[209] "A white-shoe law firm or an exclusive prep school will lean far more on recommendations and face-to-face interviews than will a fast-food chain or cash-strapped urban school district."[210] "The

---

privileged," she writes, "are processed more by people, the masses by machines."[211]

To further illustrate, consider O'Neil's story of a young man named Kyle Behm.[212] When he was a college student, Behm took some time off from school to seek treatment for bipolar disorder. When he returned to another school to finish his degree, he discovered that he kept being turned down, over and over again for job interviews. Why? He discovered that his inability to get an interview stemmed from some answers to a personality test that had been administered prior to his interviews that graded him for a series of social considerations—agreeableness, conscientiousness, neuroticism, and other qualities.[213] As Kyle's father, a lawyer, soon discovered, a number of corporations rely on the use of those tests, potentially running afoul of the Americans with Disabilities Act, which protects individuals with mental disabilities.[214]

These issues can also be exacerbated by the failure to recalibrate models, which can lead to dangerously outdated results and predictions perpetuating the continued social construction of stereotypes. O'Neil usefully contrasts this situation to how data is used by high-profile sports teams, who are constantly recalibrating and redrawing their models to ensure accuracy. For example, she explains, if the Los Angeles Lakers do not select a player because the player's data suggests that he will not be a star at scoring, and then he subsequently surpasses their expectations, the Lakers can return to their model to see how it might be improved. But in contrast, consider someone like Kyle Behm. If he finds a job somewhere and becomes a stellar employee, it is unlikely that any of the corporations that rejected him will ever know or care to return to recalibrate their model. The reason? The stakes, according to O'Neil. Individuals on basketball teams are potentially worth millions; minimum-wage employees, to the corporation making these decisions, are not worth nearly as much.[215] Unless something goes completely amiss in the workplace, the company has little reason to recalibrate its model, since the model is "doing its job—even if it misses out on potential stars."[216]  "The company may be satisfied with the status quo," O'Neil explains, "but the victims of its automatic systems suffer."[217]

---

211. *Id.*; *see also* Virginia Eubanks, Automating Inequality: How High-Tech Tools Profile, Police, and Punish the Poor (2018); Rana Foroohar, *This Mathematician Says Big Data Is Causing a 'Silent Financial Crisis'*, Time (Aug. 29, 2016), http://time.com/ 4471451/cathy-oneil-math-destruction [https://perma.cc/XW5X-7KQ5] (quoting O'Neil); *Want to Predict the Future of Surveillance? Ask Poor Communities.*, Am. Prospect (Jan. 15, 2014), http://prospect.org/article/want-predict-future-surveillance-ask-poor-communities [https://perma.cc/B8F6-DHSM].

212. Cathy O'Neil, *How Algorithms Rule Our Working Lives*, Guardian (Sept. 1, 2016), http://www.theguardian.com/science/2016/sep/01/how-algorithms-rule-our-working-lives [https://perma.cc/HJ6P-VUUN].

213. *Id.*

214. *Id.*

215. *Id.*

## A.    Surveillance, Targeting, and Stereotyping

Consider how models interface with consumers through behavioral targeting of advertising.  Here, machine learning algorithms learn from existing inputs and then limit the range of options a consumer sees.  Since websites often rely on predictive algorithms to analyze people's online activities—web surfing, online purchases, social media activities, public records, store loyalty programs, and the like—they can create profiles based on user behavior, and predict a host of identity characteristics that marketers can then use to decide the listings that a user sees online.[218]  Or their models might assign lower rankings to individuals based on their race or gender, rendering them less relevant to potential employers, limiting the scope of opportunities that they see online.[219]  Behavioral marketing has advanced to the point where advertisers can discover what motivates a given consumer and dynamically construct a particularized pitch based on the person's cognitive style (noting, for example, whether a person is impulsive or deliberative—a phenomenon that feeds into what Ryan Calo and others call "persuasion profiling").[220]

As the ACLU has argued, this sort of behavioral targeting can lead to actions that violate basic civil rights protections under the Fair Housing Act or Title VII of the Civil Rights Act.[221]  The recent lawsuit, discussed in the Introduction, is but one example of these possibilities.  But far more often, these instances reflect a kind of bias that, while demonstrative of structural discrimination, is also difficult for the law to plainly address.  For instance, researchers at Carnegie Mellon University found that Google tends to show women ads for lower paying jobs.[222]   Although the researchers never conclusively proved why, they speculated that Google's algorithms learned from the existing inequalities in society: Women are more accustomed to and associated with lower paying work, thus they tend to click on ads for lower paying jobs.[223]   The machine learning algorithm extrapolated from that behavior and continued the pattern.[224]

---

216.  *Id.*

217.  *Id.*

218.  *See* Esha Bhandari & Rachel Goodman, *ACLU Challenges Computer Crimes Law That Is Thwarting Research on Discrimination Online*, ACLU: FREE FUTURE (June 29, 2016, 10:00 AM),   http://www.aclu.org/blog/free-future/aclu-challenges-computer-crimes-law-thwarting-research-discrimination-online [https://perma.cc/UR7Y-UKJ3].

219.  *See id.*

220.  Ryan Calo, *Digital Market Manipulation*, 82 GEO. WASH. L. REV. 995, 1017 (2014); *cf.* Amit Datta et al., *Discrimination in Online Advertising: A Multidisciplinary Inquiry*, 81 PROC. MACHINE LEARNING RES. 1 (2018) (exploring interaction between tailored marketing and discrimination).

221.  Bhandari & Goodman, *supra* note 218.

In another illustrative experiment, an ad for STEM jobs that was supposed to be gender neutral in delivery appeared 20 percent more times to men than to women.[225]   The reason, the researchers postulated, was not because men were more likely to click onto the ad (women actually were more likely to click and view the ad).[226]   Rather, women aged 25–34 were the most valued, and hence, most expensive demographic to which to display ads.[227]   Here, even when there is no intent to discriminate against viewers, market principles that disproportionally value female audiences can lead to a world where AI facilitates a disparate impact on particular groups.

Scholar Karen Yeung has usefully argued that big data's harvesting of personal digital data is particularly troubling because of the ways in which advertisers use that data to nudge the user to reach decisions that accord with their commercial goals.[228]   According to Richard Thaler and Cass Sunstein, a nudge is "any aspect of choice architecture that alters people's behavior in a predictable way without forbidding any options or significantly changing their economic incentives."[229]   These modes of personalization may seem unobtrusive and subtle, but they are also incredibly powerful at the same time, echoes Yeung.[230] Since so much decisionmaking often occurs subconsciously, passively, and unreflectively—instead of through conscious deliberation—scholars have documented how even subtle changes can have a dramatic influence on decisionmaking behavior.[231] As Ryan Calo has similarly argued, these practices can rise to the level of market manipulation, because they essentially remake the consumer into what he describes as a "mediated consumer," who "approaches the marketplace through technology designed by someone else."[232]

---

222. Amit Datta, Michael Carl Tschantz & Anupam Datta, *Automated Experiments on Ad Privacy Settings: A Tale of Opacity, Choice, and Discrimination*, 2015 PROC. ON PRIVACY ENHANCING TECH. 92, 92.

223. *Id.* at 92, 105 ("Even if we could, the discrimination might have resulted unintentionally from algorithms optimizing click-through rates or other metrics free of bigotry. Given the pervasive structural nature of gender discrimination in society at large, blaming one party may ignore context and correlations that make avoiding such discrimination difficult."); *see also* Gourarie, *supra* note 43; Samuel Gibbs, *Women Less Likely to be Shown Ads for High-Paid Jobs on Google, Study Shows*, GUARDIAN (July 8, 2015, 6:29 AM), https://www.theguardian.com/technology/2015/jul/08/women-less-likely-ads-high-paid-jobs-google-study [https://perma.cc/GZ93-GNNE] (discussing the study).

224. Datta, Tschantz & Datta, *supra* note 222; Gourarie, *supra* note 43 (discussing the study).

225. Anja Lambrecht & Catherine Tucker, *Algorithmic Bias? An Empirical Study Into Apparent Gender-Based Discrimination in the Display of STEM Career Ads*, 65 MGMT. SCI (forthcoming 2019) (manuscript at 2–3) (on file with author).

226. *Id.* at 3.

227. *See id.* at 26–27.

228. *See* Karen Yeung, *'Hypernudge': Big Data as a Mode of Regulation by Design*, 20 INFO., COMM. & SOC'Y 118 (2017) (exploring the role of big data in nudging); *see also* RICHARD H. THALER & CASS R. SUNSTEIN, NUDGE: IMPROVING DECISIONS ABOUT HEALTH, WEALTH, AND HAPPINESS (2008).

229. *See* Yeung, *supra* note 228, at 120 (quoting THALER & SUNSTEIN, *supra* note 228, at 6).

Another type of issue stems from situations where particular searches correlate with other characteristics, leading to results that inaccurately suggest causal associations between a protected category and an undesirable activity. For example, Harvard researcher Latanya Sweeney found disparities in the ads Google showed alongside searches for the names Latanya and Latisha, which triggered ads for arrest records, while the names Kristen and Jill did not (even when there were arrest records associated with the names). Over 2000 names that were more likely to be associated with African American or white individuals bore out this pattern.[233] Why?[234] One possible explanation, she posited, is that people may be more likely to click on an ad for an arrest record after searching for a black name, perhaps to confirm their biases, making ads for arrest records more likely to appear for searches of those names in the future.[235] Here, the results flow "not from the racist intentions of the . . . algorithms' programmers, but from the algorithms' natural operation in the real world."[236]

As Frank Pasquale observed, the programmer might construe her role as largely agnostic, casting the search engine as a sort of "cultural voting machine, merely registering, rather than creating, perceptions."[237] However, these results, which stem from incomplete inputs in data, can produce skewed real-life perceptions of reality. For example, only 11 percent of the top 100 "CEO" image search results from Google included women, even though 27 percent of CEOs in the United States are women.[238] These biased results might at first glance, seem minor. However, over time, they can congeal into lasting, inaccurate predictions of what reality looks like, affecting "everything from personal preconceptions to hiring practices."[239] In other words, there is a risk of creating further feedback loops by presenting stereotypical information to the public without any accompanying critique. Jeremy Kun describes another, even more troubling, example. Insert "transgenders are" in Google, he suggests.[240] The results are an astonishing list of hateful descriptions—described as "freaks," "gross," "sick," "wrong," and "crazy."[241] But that is what AI reveals—a trend towards hateful autocompletes. Those autocompletes can actually wind up feeding into stereotypes, leading to further examples of biased social constructions as a result of incomplete information.

---

230. *Id.* at 119.
231. *Id.* at 120. "Nudges" can operate through automated decisionmaking techniques that might act dynamically in the following ways: (1) by refining an individual's choice environment based on his or her data profile, (2) providing feedback to the choice architect, to be stored and repurposed, and (3) by monitoring and refining an individual's choice environment given broader population-wide trends gleaned from data surveillance and analysis. *Id.* at 122.
232. Calo, *supra* note 220, at 1002.
233. Latanya Sweeney, *Discrimination in Online Ad Delivery*, 11 Comms. ACM 44, 46–47, 50–51 (2013).
234. *Id.* at 52.
235. *Id.*
236. Anupam Chander, *The Racist Algorithm?*, 115 Mich. L. Rev. 1023, 1037 (2017).
237. *Id.* (quoting Pasquale, *supra* note 18, at 39).

Beyond issues of discrimination, algorithms can raise privacy concerns as well. Consider the famous example of the Target algorithm that predicted a woman's pregnancy even before her family knew that she was pregnant, and then used this knowledge for marketing purposes.[242] As Kate Crawford and Jason Schultz have observed, Target "did not collect the information from any first or third party," and therefore was not required to notify its customers that it was using this information in the same way that other collection protocols require.[243] In such contexts, they point out that the concept of differential privacy is severely limited, since it is "impossible" to tell when or where to assemble such protections around the inputs provided by the end user.[244] As

---

238. Matthew Kay, Cynthia Matuszek, & Sean A. Munson, *Unequal Representation and Gender Stereotypes in Image Search Results for Occupations*, *in* CHI 2015: PROCEEDINGS OF THE 33RD ANNUAL CHI CONFERENCE ON HUMAN FACTORS IN COMPUTING SYSTEMS 3819 (2015); Chloe Albanesius, *When You Google Image Search 'CEO,' the First Woman Is . . .* , PC MAG. (Apr. 12, 2015, 6:35 PM), http://www.pcmag.com/article2/0,2817,2481270,00.asp [https://perma.cc/K9XZ-REPZ] (citing report); *see also* T.C. Sottek, *Google Search Thinks the Most Important Female CEO Is Barbie*, VERGE (Apr. 9, 2015, 3:28 PM), https://www.theverge.com/tldr/2015/4/9/8378745/i-see-white-people [https://perma.cc/X3S3-8XGC] (observing that the first woman appearing in a google image search for CEO is an image of a Barbie doll).

239. Albanesius, *supra* note 238.

240. Kun, *supra* note 16. Also see Safiya Noble, author of ALGORITHMS OF OPPRESSION (2018), discussing a similar result when inputting "black girls." Safiya Noble, *Google Has a Striking History of Bias Against Black Girls*, TIME (Mar. 26 2018), http://time.com/5209144/google-search-engine-algorithm-bias-racism.

241. Kun, *supra* note 16.

242. *See* Kate Crawford & Jason Schultz, *Big Data and Due Process: Toward a Framework to Redress Predictive Privacy Harms*, 55 B.C. L. REV. 93, 94 (2014).

243. *Id.* at 98.

244. *Id.* at 99.

Crawford and Schultz ask, "[w]hen a pregnant teenager is shopping for vitamins, could she predict that any particular visit or purchase would trigger a retailer's algorithms to flag her as a pregnant customer? And at what point would it have been appropriate to give notice and request her consent?"[245]

## B.    Price Discrimination and Inequality

In 2012, an Atlanta man returned from his honeymoon to find that his credit limit had been lowered from $10,800 to $3,800. He had not defaulted on any debts. Nothing in his credit report had changed. American Express cited aggregate data. A letter from the company told him, "[o]ther customers who have used their card at establishments where you recently shopped have a poor repayment history with American Express."[246] Similarly, one credit card company settled FTC allegations that it failed to disclose its practice of rating consumers as having a greater credit risk "because they used their cards to pay for marriage counseling, therapy, or tire-repair services," based on its experiences with other consumers and their repayment histories.[247]

As these examples suggest, the intersection of machine learning and automated decisionmaking can make particular groups materially worse off, by charging higher prices or interest rates, or excluding them entirely. Unlike false information on a credit report, no federal law allows consumers to challenge the generalizations that algorithms make about them based on other data from social media or search engines.[248]

Yet quantative models can also help companies price discriminate against certain consumers, charging more for the same goods. In the case of health insurance, and a host of other industries, this practice may lead to higher premiums based on potentially irrelevant characteristics.[249] With the help of machine learning, companies have begun to price discriminate in the third-degree, using features besides ability and willingness to set prices.[250] Indeed, if the algorithm is poorly written and the data is biased, these factors can become a proxy for elasticity of demand. When elasticity of demand is low, monopolies flourish, raise prices for all consumers, and exclude less affluent consumers from the market.[251]

---

245.  *Id.*

246.  Andrews, *supra* note 150.

247.  Citron & Pasquale, *supra* note 27, at 4; Stipulated Order for Permanent Injunction and Other Equitable Relief, FTC v. CompuCredit Corp., No. 1:08-CV-1976-BBM-RGV, 2008 WL 8762850 (N.D. Ga. Dec. 19, 2008), http://www.ftc.gov/sites/default/files/documents/cases/2008/12/081219compucreditstiporder.pdf [https://perma.cc/CBU6-867U].

248.  *See, e.g.*, Citron & Pasquale, *supra* note 27, at 4–5; *see also* Nat'l Consumer Law Ctr., Comments to the Federal Trade Commission, Big Data: A Tool for Inclusion or Exclusion?   Workshop,   Project   No.   P145406   (Aug.   15,   2014), http://www.ftc.gov/system/files/documents/public_comments/2014/08/00018-92374.pdf [https://perma.cc/RM4K-DJYV] (citing Persis Yu et al., Nat'l Consumer Law Ctr., Big Data: A Big Disappointment for Scoring Consumer Credit Risk (2014)).

249.  *See* Crawford et al., *supra* note 63, at 6–7. *See* Exec. Office of the President, Big Data and Differential Pricing 17 (2015), http://obamawhitehouse.archives.gov/sites/default/files/whitehouse_files/docs/Big_Data_Report_Nonembargo_v2.pdf

The internet is already rife with examples of algorithms artificially inflating prices in the name of optimization. Amazon's pricing algorithm once set the price of Peter Lawrence's book, *The Making of a Fly*, at $23,698,655.93 for all consumers.[252] The inefficiency of that price is obvious. However, other companies have muddied the line between price discrimination and price optimization in much subtler and ultimately effective ways, raising concerns that optimization enables insurers to raise premiums on customers who may not aggressively shop around for better rates (or are perceived to avoid doing so).[253] In one example involving auto insurance, consumers who were likely to compare prices could receive up to a 90 percent discount, while others could see their premiums increase by up to 800 percent.[254] In another example of this

---

[https://perma.cc/Q4ST-8UXU] ("If price-sensitive customers also tend to be less experienced, or less knowledgeable about potential pitfalls, they might more readily accept offers that appear fine on the surface but are actually full of hidden charges."). *See generally* Dirk Bergemann et al., *The Limits of Price Discrimination*, 105 Am. Econ. Rev. 921 (2015) (analyzing price discrimination and its implications).

250. *See* Bergemann, *supra* note 249, at 926–27.

251. *See generally* Ariel Ezrachi & Maurice E. Stucke, Virtual Competition: The Promise and Perils of the Algorithm-Driven Economy (2016) (exploring implications of algorithms for market competition).

252. Michael Eisen, *Amazon's $23,698,655.93 Book About Flies*, It is Not Junk (Apr. 22, 2011), http://www.michaeleisen.org/blog/?p=358 [https://perma.cc/4LAS-2DE5].

253. Julia Angwin et al., *Minority Neighborhoods Pay Higher Car Insurance Premiums Than White Areas With the Same Risk*, ProPublica (Apr. 5, 2017), http://www.propublica.org/article/minority-neighborhoods-higher-car-insurance-premiums-white-areas-same-risk [https://perma.cc/9RWT-XGBB].

254. *Watchdog: Allstate Auto Insurance Pricing Scheme Is Unfair*, AOL (Dec. 16, 2014, 1:02 PM) https://www.aol.com/article/finance/2014/12/16/allstate-auto-insurance-pricing-scheme-unfair/21117081 [https://perma.cc/898E-9T5U]. In 2015, Consumer Reports detailed a number of factors used by insurance companies, including "price optimization" strategies that relied on personal data and statistical models to predict a person's likelihood to comparison shop. Fredrick Kunkle, *Auto Insurance Rates Have Skyrocketed—And in Ways That Are Wildly Unfair*, Wash. Post (Feb. 7, 2018), https://www.washingtonpost.com/news/tripping/wp/2018/02/07/auto-insurance-rates-have-skyrocketed-and-in-ways-that-are-wildly-unfair/?noredirect=on&utm_term=.040bec7b1522 [https://perma.cc/BCC6-ZJV3] (mentioning 2015 report and its impact on auto insurance); *cf.* Tracy Samilton, *Being a Loyal Auto Insurance Customer Can Cost You*, NPR (May 8, 2015, 3:51 AM), http://www.npr.org/2015/05/08/403598235/being-a-loyal-auto-insurance-customer-can-cost-you. Courts and insurance regulators in a number of states trend, Orbitz showed more expensive hotels to Mac users, apparently believing that the operating system was a proxy for affluence.[255]  Orbitz showed less expensive hotels to users on mobile devices, which minority groups tend to use at higher rates.[256]

Many of these issues, as I have suggested, escape legal detection because they are not clearly illegal.  And even if they were, the opacity of algorithmic design and decisionmaking makes many of these issues difficult to detect.  Moreover, the absence of privacy protections further exacerbates the problem.  For example, one study of over 80,000 health-related web pages discovered that over 90 percent of those sites shared user information with outside third parties.[257] Visitors to the Centers for Disease Control sites, for example, had their browsing information shared with Google, Facebook, Pinterest, and Twitter, often without their knowledge or consent, and often outside of the reach of privacy statutes like the Health Insurance Portability and Accountability Act.[258]

In such cases, even if advertisers do not know the name or identity of the person searching for information, the data can still be aggregated to "paint a revealing portrait" of the person.[259]  For example, Facebook could link health-related web browsing to an identifiable person, leading to some measurable risk that such information could be misused by other companies who seek to profit from it.[260]  Expert Tim Libert offers the example of data broker MedBase200, which sold lists of individuals under such categories as "rape sufferers," "domestic abuse victims," or "HIV/AIDS patients."[261]  While it is unclear how

---

have begun to crack down on the use of extraneous factors in insurance pricing.  *See* Stevenson v. Allstate Ins. Co., No. 15-CV-04788-YGR, 2016 WL 1056137, at *2 (N.D. Cal. Mar. 17, 2016) ("Earnix software allows Allstate to account for ED [elasticity of demand] in its rating factors submitted for approval, without disclosing to [the California Department of Insurance] that it is considering ED when compiling its class plan." (citation omitted)).

255.  Orbitz claimed that the algorithm responded to the fact that Mac users already spent as much as 30 percent more a night on hotels before Orbitz implemented the pricing algorithm.  Dana Mattioli, *On Orbitz, Mac Users Steered to Pricier Hotels*, Wall Street J. (Aug. 23, 2012, 6:07 PM), http://www.wsj.com/articles/SB1000142405270230445860457488822667325882.

256.  *See* Jennifer Valentino-DeVries et al, *Websites Vary Prices, Deals Based on Users' Information*, Wall Street J. (Dec. 24, 2012), https://www.wsj.com/articles/SB10001424127887323777204578189391813881534.

257.  Tim Libert, *Health Privacy Online: Patients at Risk*, *in* Open Tech. Inst. & New Am., Data and Discrimination: Collected Essays 11, 12 (Seeta Peña Gangadharan et al. eds., 2014), http://na-production.s3.amazonaws.com/documents/data-and-discrimination.pdf [https://perma.cc/9GGR-ZXDC].

258.  *Id.* at 12–13.

259.  *Id.* at 13.

260.  *Id.*

261.  *Id.*

MedBase200 obtained such data, the risk of data brokers purchasing such information and misusing it is apparent.[262] These situations raise the risks of user identification, price discrimination, or other forms of mistreatment. Given that online advertisers often categorize information into target and nontarget users, there is a significant risk that a user may be discriminated against based on his or her web-browsing activities.[263] Libert offers the observation that, since over 60 percent of bankruptcies are medically related, companies could potentially target certain individuals for more favorable discounts than those who fall into nontarget categories in the absence of smarter, and more specific regulation.[264]

The *New York Times* wrote about a crop of banking startups that used inferences from big data to identify populations that might be ignored by traditional lenders—creditworthy, but not necessarily with the assets to make a large down payment on a mortgage.[265] The company used factors like whether applicants type in ALL CAPS, as well as how much time they spent reading terms and conditions, to determine creditworthiness.[266] While we might have suspicions about the habits of people who write in ALL CAPS or blithely ignore terms and conditions,[267] there is no empirical reason to believe they are less creditworthy than their less emphatic counterparts. Another company used a large dataset to determine that "people who fill out online job applications using browsers that did not come with the computer . . . but had to be deliberately installed (like Firefox or Google's Chrome) perform better and change jobs less often."[268]

## III.    RETHINKING CIVIL RIGHTS THROUGH PRIVATE ACCOUNTABILITY

These applications of data analytics require us to think broadly about how to address inequality and discrimination in a digital age. But some of this requires a fundamental rethinking of civil rights protections. Lawyers, accustomed to constitutional concepts like due process and privacy, struggle to map these concepts onto private corporations' novel practices. Some of these practices can interface with public institutions, like traditional law enforcement, creating a greater set of possibilities for accountability through the application of constitutional principles. Others can bring private causes of action. Stretching these lofty protections beyond the state into the private sphere, however, can present challenges.[269]

---

262.  *Id.*
263.  *See id.* at 14.
264.  *Id.*
265.  *See* Steve Lohr, *Banking Start-Ups Adopt New Tools for Lending*, N.Y. TIMES (Jan. 18, 2015), http://www.nytimes.com/2015/01/19/technology/banking-start-ups-adopt-new-tools-for-lending.html.
266.  *Id.*
267.  One study, however, suggests that 98 percent of people do not closely read all terms of service. *See* Shankar Vedantam, *Do You Read Terms of Service Contracts? Not Many Do, Research Shows*, NPR (Aug. 23, 2016, 5:06 AM), http://www.npr.org/2016/08/23/491024846/do-you-read-terms-of-service-contracts-not-many-do-research-shows.
268.  *Robot Recruiters: How Software Helps Firms Hire Workers More Efficiently*, ECONOMIST (Apr. 6, 2013), http://www.economist.com/news/business/21575820-how-software-helps-firms-hire-workers-more-efficiently-robot-recruiters [https://perma.cc/W9GU-DDFK].

But there is another, deeper reason for why this age of machine learning is so transformative, and that is because it forces us to reevaluate our entire spectrum of civil rights in the process. Like the civil rights era that came before it, AI is implicated within a vast array of decisions that come, not from the government, but from the private sector, even if many of them implicate civil rights in the process. For example, the right to be considered for employment, free from consideration of one's disability—the right at issue in the Kyle Behm case just discussed—directly correlates to the right to work. Similarly, the right to an education, the right to vote, the right to make contracts, the right to travel, the right to get insurance, and the right to receive information, among others, are all at issue when an algorithm makes its (private) decisions about who does and who does not receive the entitlement and the conditions attached to it. Those decisions are not always subject to public oversight. And even more problematically, they may be shielded from view, due to trade secrecy and systemic opacity.

The Obama-era White House was far from blind to the risks and benefits of big data. It concluded that "big data analytics have the potential to eclipse longstanding civil rights protections in how personal information is used in housing, credit, employment, health, education, and the marketplace."[270] Previously, the administration recommended the development of algorithmic auditing and fairness considerations;[271] it remains to be seen what the current administration will do, if anything.

In 2013, the United States Consumer Financial Protection Bureau and the Department of Justice reached an $80 million settlement with Ally Financial Inc., an auto lender that allegedly added significant "dealer markups" to minority

---

borrowers.[272] The markups allegedly led to African American borrowers paying, on average, nearly $300 more than their white counterparts, and Hispanic borrowers more than $200.[273] The government figured this out after using an algorithm, known as the Bayesian Improved Surname Geocoding (BISG) that estimated the probability of a borrower being a minority by using a person's last name and location.[274] Admittedly, the algorithm is imperfect, leading to a number of nonminority positives,[275] but it does provide a helpful tool for uncovering hidden biases.

Yet as our current laws stand, there is little that has been done to address the problem of algorithmic bias. First, our existing frameworks for regulating privacy and due process cannot account for the sheer complexity and numerosity of cases of algorithmic discrimination. Second, our existing statutory and constitutional schemes are poorly crafted to address issues of private, algorithmic discrimination. In part because of these reasons, private companies are often able to evade statutory and constitutional obligations that the government is required to follow. Third, because of the dominance of private industry, and the concomitant paucity of information privacy and due process protections, individuals can be governed by biased decisions and never realize it, or they may be foreclosed from discovering bias altogether due to the lack of transparency. These situations, in turn, limit the law's ability to address the problem of bias. Elizabeth Joh, for example, has written extensively about how surveillance technology developed by private companies—including big data programs—has exercised undue influence over policing, overriding principles of transparency and accountability which normally govern police departments, distorting the reach of the Fourth Amendment.[276]

## A.    The Paucity of the Principle of Nondiscrimination

As Danah Boyd and others have noted, the notion of a protected class is also a fuzzy category in practice. "The notion of a protected class remains a fundamental legal concept, but as individuals increasingly face technologically mediated discrimination based on their positions within networks, it may be incomplete."[277] Since the range of potential inputs for discrimination is so much broader, it is "increasingly hard to understand what factors are inputted or inferred in complex algorithms that seek to distribute limited resources."[278]

---

272. *CFPB and DOJ Order Ally to Pay $80 Million to Consumers Harmed by Discriminatory Auto Loan Pricing*, CFPB: NEWSROOM (Dec. 20, 2013), http://www.consumerfinance.gov/ about-us/newsroom/cfpb-and-doj-order-ally-to-pay-80-million-to-consumers-harmed-by-discriminatory-auto-loan-pricing [https://perma.cc/HCZ7-A3SV].

273. AnnaMaria Andriotis & Rachel Louise Ensign, *U.S. Government Uses Race Test for $80 Million in Payments*, WALL STREET J. (Oct. 29, 2015), http://www.wsj.com/articles/u-s-uses-race-test-to-decide-who-to-pay-in-ally-auto-loan-pact-1446111002.

274. *Id.*

275. *Id.*

276. Joh, *supra* note 269, at 103.

As Barocas and Selbst have insightfully noted, data mining techniques force a central confrontation between two central principles that underscore antidiscrimination law: anticlassification and antisubordination.[279] Anticlassification principles suggest that the very act of classification risks unfairness for individuals in protected groups because decisionmakers may rest their judgments on inappropriate perceptions. In contrast, antisubordination theory aims to remedy unequal treatment as a matter of substance (as opposed to procedure), pointing out that the central goal of antidiscrimination law should be to eliminate any status-based distinctions between protected and unprotected categories.[280] In order for the law to address the risk of discrimination in an algorithmic context, it is necessary for legislatures to commit to antisubordination principles in a way that they have not been able to do, since courts are exercising more and more scrutiny over substantive remediation.[281] If these remedies remain both politically and constitutionally infeasible, then antidiscrimination principles may never be able to fully address discrimination in data mining techniques.[282]

In *Ricci v. DeStefano*, a case in which the City of New Haven refused to certify a promotion exam on the basis that it would have produced a disparate impact, we see a powerful enactment of these concerns.[283] Even though the city's refusal constituted a facially neutral effort to correct for a perceived disparate impact regarding race, the U.S. Supreme Court concluded that the City's refusal constituted a race-conscious remedy that comprised disparate treatment of the white firefighters who might have been promoted based on the results of the exam.[284] Disparate treatment, the Court concluded, cannot serve as a remedy for disparate impact, without a strong showing that the initial results would lead to liability for actual disparate treatment.[285]

---

277. Danah Boyd et al., *The Networked Nature of Algorithmic Discrimination*, *in* DATA AND DISCRIMINATION: COLLECTED ESSAYS, *supra* note 257, at 53, 56.
278. *Id.*
279. Barocas & Selbst, *supra* note 60, at 723.
280. *Id.*
281. *Id.*
282. *Id.*
283. *Id.* at 724 (discussing Ricci v. DeStefano, 557 U.S. 557 (2009)).
284. *Id.* at 724–25.
285. *Id.* at 725.

Taking *Ricci* at its word, Borocas and Selbst suggest that legislative attempts to require certain types of remedial action in discriminatory data mining may run afoul of the existing nexus that bars disparate treatment as a solution for problems regarding disparate impact.[286] Even if Congress amended Title VII to force employers to make their training data and models auditable to focus on an algorithm's discriminatory potential, any solution would necessarily require the employer to first consider membership in a protected class, thus raising the spectre of a race-conscious remedy.[287] Although the authors note that it is possible to explore the potential of discriminatory impact at the test design stage under *Ricci*, they argue that "[a]fter an employer begins to use the model to make hiring decisions, only a 'strong basis in evidence' that the employer will be successfully sued for disparate impact will permit corrective action."[288] This high threshold makes the opportunities for such corrective action quite limited, since, as the authors point out, "disparate impact will only be discovered after an employer faces complaints," and then forces an investigation.[289]

As this discussion illustrates, our traditional civil rights principles, particularly in the world of Title VII, map unevenly onto a world that facilitates algorithmic discrimination.[290] Further difficulties of finding proof of both a discriminatory intent and impact abound, since most data mining practices would not automatically generate liability under Title VII.[291] Part of this is attributable to the way in which Title VII constructs standards of proof. Even when data mining results in a discriminatory impact, as Barocas and Selbst explain, the law is constructed to balance the protection of legitimate business judgments with "preventing 'artificial, arbitrary, and unnecessary' discrimination."[292] If the two happen to conflict, they conclude, "a tie goes to the employer."[293]

But there is also another, constitutional barrier towards equality. Since procedural remedies may not be able to solve many of the problems associated with big data discrimination, it may often be necessary to rebalance variables, reweighting results in order to compensate for discriminatory outcomes.[294] On this, Anupam Chander has suggested a number of such possibilities in a recent piece.[295] However, it is important to note that any rebalancing effort may not survive our current constitutional climate because these efforts, at least in the race-based context, raise constitutional concerns due to the specter of affirmative action.

---

286. *Id.*
287. *Id.*
288. *Id.* at 725–26 (quoting *Ricci*, 557 U.S. at 585).
289. *Id.* at 726.
290. For a related perspective on *Ricci* and the role of audits, see Pauline T. Kim, *Auditing Algorithms for Discrimination*, 166 U. PA. L. REV. ONLINE 189 (2017) (discussing desirability and practicality of audits to detect discrimination).
291. *See* Barocas & Selbst, *supra* note 60, at 726.
292. *Id.* at 711 (quoting Griggs v. Duke Power Co., 401 U.S. 424, 431 (1971)).
293. *Id.* at 712.
294. *Id.* at 715.

## B. The Paucity of Privacy Protections

Other existing normative commitments to civil rights can also be just as inadequate when we try to apply them to algorithmic accountability. Take informational privacy as one example.[296] Beyond just the absence of granular, statutory language protecting informational privacy concerns, as mentioned throughout this Article in the context of health, there are other major obstacles to informational privacy's ability to address algorithmic discrimination.

One obstacle is simple awareness and lack of notice. Privacy, Oscar Gandy writes, is not going to solve the problem of disparate impact where the algorithm is concerned.[297] Most of the time, people who might be discriminated against by a systemic issue might not even know that they are being discriminated against at all.[298]

But there is a deeper reason for the absence of greater regulation in the United States. As Paul Schwartz and Karl-Nikolaus Peifer have usefully explained, the U.S. and the EU systems diverge substantially in their approaches to privacy.[299]  The EU structures its system of privacy regulation through the lens of a fundamental set of rights that address data protection, largely through a rights-based set of entitlements.[300]  Within this model, the EU privileges the individual through a discourse that relies on the language of constitutional rights anchored by the values of dignity, personality, and self-determination, drawn in no small part from the European Convention of Human Rights and the Charter of Fundamental Rights, both of which have led to an explicit right to data protection encircling both the government and private parties.[301]  Although the free flow of information is also an important value in this system, it matters less, according to Schwartz and Peifer, than the individual right to dignity, privacy, and data protection.[302]

---

295. Chander, *supra* note 236, at 1041–42 (detailing the potential for modeling remedies to algorithmic discrimination on affirmative action).

296. *See* Cynthia Dwork & Aaron Roth, *The Algorithmic Foundations of Differential Privacy*, 9 FOUND. & TRENDS THEORETICAL COMPUTER SCI. 211 (2014), http://www.cis.upenn.edu/~aaroth/Papers/privacybook.pdf [https://perma.cc/7NA5-6Y8R] (arguing for a more robust definition of privacy through algorithmic analysis).

297. Oscar H. Gandy, Jr., *Engaging Rational Discrimination: Exploring Reasons for Placing Regulatory Constraints on Decision Support Systems*, 12 J. ETHICS & INFO. TECH. 29, 39–40 (2010). For other excellent treatments on privacy, see JULIE E. COHEN, CONFIGURING THE NETWORKED SELF: LAW, CODE, AND THE PLAY OF EVERYDAY PRACTICE (2012); HELEN NISSENBAUM, PRIVACY IN CONTEXT: TECHNOLOGY, POLICY, AND THE INTEGRITY OF SOCIAL LIFE (2010); JEFFREY ROSEN, THE UNWANTED GAZE: THE DESTRUCTION OF PRIVACY IN AMERICA (Vintage Books rev. ed. 2001); PAUL M. SCHWARTZ, THE CTR. FOR INFO. POLICY LEADERSHIP, DATA PROTECTION LAW AND THE ETHICAL USE OF ANALYTICS (2010), http://iapp.org/media/pdf/knowledge_center/Ethical_Underpinnings_of_Analytics.pdf [https://perma.cc/A2FQ-PG9S]; DANIEL J. SOLOVE, NOTHING TO HIDE: THE FALSE TRADEOFF BETWEEN PRIVACY AND SECURITY (Yale Univ. Press rev. ed. 2013); DANIEL J. SOLOVE, UNDERSTANDING PRIVACY (Harv. Univ. Press rev. ed. 2010); Paul Ohm, *Broken Promises of Privacy: Responding to the Surprising Failure of Anonymization*, 57 UCLA L. REV. 1701 (2010); Jane Yakowitz, *Tragedy of the Data Commons*, 25 HARV. J.L. & TECH. 1 (2011).

298. *See, e.g.*, Angwin et al., *supra* note 149 (discussing how Facebook may have more information on its users than we realize).

In contrast, the United States employs a more market-driven structure, viewing the individual through a consumerist lens that focuses on the individual as a "privacy consumer,"—a "trader of a commodity, namely her personal data."[303]  Here, the focus on privacy is framed as a matter of "bilateral self-interest," leading to a focus on "policing fairness in exchanges of personal data."[304]  The Constitution, in this framework, does not govern horizontal, private-to-private exchanges between individuals, nor does it "oblige the government to take positive steps to create conditions to allow for the existence of fundamental rights."[305]  Although there are some sources of protection from the Fourth Amendment and the Due Process Clause of the Fourteenth Amendment, those map unevenly onto the concerns of information privacy.[306]

Consider the Fourth Amendment as an example.  As Schwartz and Peifer explain, since the Amendment is concerned with the reasonableness of searches and seizures, it fails to govern information that is already held by government databases, as well as situations where a third party (like a bank) hands over personal information to the government.[307]  Although the Supreme Court recognized a general right to information privacy in 1977 when it decided *Whalen v. Roe*,[308] its progeny suggests a general level of uncertainty regarding the

---

contours of such a right.[309]  Unlike in the EU, in the United States, there is no analogous right to data protection.[310]  As Schwartz and Peifer observe, this is partly a result of the uncertainty in the United States about whether a variety of information processing practices constitute evidence of sufficient harm to warrant a legal remedy.[311]  Instead, privacy protections comprise a patchwork of federal and state statutes and regulations that have been enacted without the broader, harmonizing protection that an omnibus law would provide.[312]  In addition, marketplace rhetoric favors laws that privilege notice and consent.[313]  Echoing some of these insights, Lior Strahilevitz has argued that the absence of prophylactic privacy laws in the United States, when coupled with attitudinal differences and public choice issues, makes subsequent privacy regulation more unlikely in the future.[314]  As a result, there is a failure of informational privacy protections to creatively address situations that seem like the benign sharing of information between data brokers and their advertisers.[315]

## C.    Automated Decisionmaking and Due Process

This lack of awareness ties directly into due process concerns.  Today, computers and algorithms are an essential part of government.[316]  As Danielle Keats Citron has noted in her foundational work on this topic, automated decisionmaking systems have become the primary decisionmakers for a host of government decisions, including Medicaid, child-support payments, airline travel, voter registration, and small business contracts.[317]  While automation dramatically lowers the cost of decisionmaking, it also raises significant due process concerns, involving lack of notice and the opportunity to challenge the decision.[318] The problem is not just that governmental decisionmaking has been delegated to private entities that design code; it is also the reverse situation, where private entities have significant power that is not regulated by the government.

---

309.  Schwartz & Peifer, *supra* note 299, at 133–34.

310.  *Id.* at 134.

311.  *See id.* at 135–36.

312.  *Id.* at 136.

313.  *Id.*

314.  Lior Jacob Strahilevitz, *Toward a Positive Theory of Privacy Law*, 126 HARV. L. REV. 2010, 2036 (2013).

315.  For example, in the STEM study mentioned earlier, the postulated reason for why more men than women were shown STEM-related ads was not the presumed differences between men and women, but rather the way that advertising priced the cost of male and female audiences of particular ages.  For the authors of this study, the interconnectedness of the data led to spillover effects that directed discriminatory decisions, thus demonstrating the need to reevaluate the role of privacy protections.  Instead of thinking about informational privacy protections as traditionally restraining particular actions, the researchers urged others to think about privacy in terms of its relationship to these spillovers instead.  Lambrecht & Tucker, *supra* note 225, at 4; *see also* Raymond et al., *supra* note 48, at 218 (discussing complex role of privacy protections).

316.  *See* Paul Schwartz, *Data Processing and Government Administration: The Failure of the American Legal Response to the Computer*, 43 HASTINGS L.J. 1321, 1322 (1992).

317.  Danielle Keats Citron, *Technological Due Process*, 85 WASH. U. L. REV. 1249, 1252 & n.12 (2008).

The European Union recently adopted due process requirements, based partly on the rights-based framework discussed above, that create procedures that enable citizens to receive and challenge explanations for automated decisions when they receive decisions based "solely on automated processing" and when the decisions "significantly affect" their lives.[319] Unfortunately, this right only affects a very small number of automated decisions, since those eligible individuals are those who received decisions that do not involve human intervention, like an automated refusal of a credit application.[320] However, the EU GDPR took effect in May 2018, representing perhaps the most prominent regulatory move in favor of greater protections for individuals.[321] It requires companies and governments to reveal an algorithm's purpose and the data it uses to make decisions, leading some to infer a right to explanation.[322]

The GDPR requires individuals to have the right to confirm whether their personal data is being processed, the purpose of the process, the source of the data, and the logic behind any automated decisionmaking.[323] Yet it is unclear if decisions made based on data about a large group with which the individual identified would also trigger notification.[324] As Selbst has observed, there is also

---

318. *Id.* at 1249.

319. *See* Parliament & Council Regulation 2016/679, art. 22, 2016 O.J. (L 119) 1.

320. *Id.*; *see also Rights Related to Automated Decision Making Including Profiling*, ICO, https://ico.org.uk/for-organisations/guide-to-the-general-data-protection-regulation-gdpr/individual-rights/rights-related-to-automated-decision-making-including-profiling [https://perma.cc/CFB2-EN2U].

321. *See GDPR Portal: Site Overview*, EU GDPR.ORG, http://www.eugdpr.org [https://perma.cc/7659-HRCT].

322. *Id.; see also* EUROPEAN COMM'N, EU DATA PROTECTION REFORM: BETTER RULES FOR EUROPEAN BUSINESSES, https://ec.europa.eu/commission/sites/beta-political/files/data-protection-factsheet-business_en.pdf [https://perma.cc/K7LW-2DH8]; Andrew D. Selbst & Solon Barocas, *The Intuitive Appeal of Explainable Machines*, 87 FORDHAM L. REV. (forthcoming 2019) (manuscript at 36).

323. *See id.*; *see also Rights Related to Automated Decision Making Including Profiling*, *supra* note 320; *Article 15, EU GDPR, "Right of Access by the Data Subject*," PRIVAZYPLAN, http://www.privacy-regulation.eu/en/article-15-right-of-access-by-the-data-subject-GDPR.htm [https://perma.cc/KES2-9MJ4].

324. *See* Sandra Wachter et al., *Why a Right to Explanation of Automated Decision-Making Does Not Exist in the General Data Protection Regulation*, 7 INT'L DATA PRIVACY L. 76, 88–89 (2017). *But see* Andrew D. Selbst & Julia Powles, *Meaningful Information and the Right to Explanation*, 7 INT'L DATA PRIVACY L. 233 (2017) some debate over the level of meaningfulness those explanations are required to demonstrate.[325]  To be meaningful, then, for Selbst and Barocas, the information must be about the logic behind the decision, enabling the subject to decide whether or not to invoke his or her private right of action under the GDPR.[326] Without a clear definition, as the above examples suggest, there is an appreciable risk that companies will explain their algorithms in the most innocuous way possible.[327]   A further obstacle involves trade secrecy.  According to some researchers, courts in Germany and Austria have interpreted similar, existing laws narrowly to allow companies to limit their explanations to avoid revealing trade secrets.[328]   And without legal intervention into private industry, comprehensive solutions cannot even begin to develop.

## IV.   REFINING OVERSIGHT FROM WITHIN

As suggested above, this Article argues that part of the problem has been our reliance on traditional civil rights principles to address the issue of algorithmic bias.  To solve this problem, we must begin at the same place that critical race scholars began decades ago: recognizing areas where the law has failed to protect the interests of nondiscrimination and equality.   As demonstrated throughout this Article, issues of informational privacy, equality, and due process have surfaced in a variety of algorithmic contexts, but existing law has remained inadequate in addressing this problem, in part due to issues surrounding detection.   Other obstacles, as I have suggested, stem from the significant information asymmetry between those who design algorithms and those who are governed by them.   A third obstacle, as demonstrated by the absence of a GDPR comparative in the United States, stems from the absence of meaningful regulation to address issues of transparency and accountability.

As I suggest below, part of the answer lies in meaningful responses from private industry in order to address the problem.  In turn, the gaping absence of regulatory oversight, particularly in the current administration, requires us to turn to two other potential avenues for greater transparency: voluntary self-regulation (discussed below) and individual actions by employees through

---

whistleblowing (discussed in Part V).[329]  Part of what this issue requires is also a fundamental rethinking of the relationship between civil rights, consumer protection, and automated decisionmaking.  Instead of looking to the government for protection from algorithmic bias, society must look elsewhere.

Today, the near-complete absence of attention from our current government suggests a much greater need to explore models of self-regulation, even though, of course, government intervention would be much more effective.  In the following Subparts, I explore the possibilities for addressing algorithmic accountability from within—both from within the industry as well as from within the company itself, discussing the possibility of codes of conduct, impact statements, and whistleblowing to address the issue of algorithmic accountability.  Of course, it is also important to recognize the fact that effective self-regulation may not always resolve the issue of algorithmic fairness.  Certainly, there are powerful arguments that can be made about the limited incentives for companies to rigorously examine the implications behind technologies that are both profitable and powerful.  Yet at the same time, the range of attention paid to self-regulation, from both private industry and from organizations within computer science, suggests that there may be some room to explore potential alternatives from within the industry.  And the explosion of AI-related organizations that focus on industry accountability gives us some optimism that the industry is aiming to address issues of transparency and accountability.[330]

## A.    Codes of Conduct

The issue of algorithmic accountability demonstrates one core aspect that is missing among computer scientists and software engineers: a concrete, user-friendly, ethical platform with which to approach decisionmaking and software design.  Indeed, one might argue that the absence of this platform has facilitated algorithmic decisionmaking without recognition of the societal effects of these decisions on the poor and other protected groups.  Consequently, restoring some modicum of ethical decisionmaking may be one part of the solution.

---

329.  Will Knight, *Biased Algorithms Are Everywhere, and No One Seems to Care*, MIT TECH. REV. (July 12, 2017), https://www.technologyreview.com/s/608248/biased-algorithms-are-everywhere-and-no-one-seems-to-care [https://perma.cc/XV7M-XSL9] ("[T]he Trump administration's lack of interest in AI—and in science generally—means there is no regulatory movement to address the problem.").

330.  See, for example, the work being done by AI Now, Partnership on Artificial Intelligence, Future of Humanity Institute, and others.

Recently, researchers at Amazon, Facebook, IBM, Microsoft, and Alphabet have been attempting to design a standard of ethics around the creation of artificial intelligence.[331] Another possible form of self-regulation involves the development of a set of ethical principles within professional organizations like the Association for the Advancement of Artificial Intelligence (AAAI) and the Association of Computing Machinery (ACM).[332] Despite these admirable efforts of self-regulation, to be truly effective, these principles must be promulgated both within the AI community, as well as distributed to a variety of other professional organizations that draw on big data—like health, financial, and government sectors.[333] They also require regulatory participation to be most effective. But even in the absence of such oversight, they are still worth serious consideration.

The ACM, for example, has established seven principles for Algorithmic Transparency and Accountability, noting the importance of: (1) awareness of possible biases in design, implementation, and use; (2) access and redress mechanisms to allow individuals to question and address adverse effects of algorithmically informed decisions; (3) accountability, ensuring that individuals are held responsible for decisions made by algorithms that they use; (4) an explanation regarding both the procedures that the algorithm follows as well as the specific decisions that are made; (5) data provenance, meaning a description of the way that the training data was collected, along with "an exploration of the potential biases induced by the human or algorithmic data-gathering process"; (6) auditability, enabling models, algorithms, data and decisions to be recorded for audit purposes; and (7) validation and testing, ensuring the use of rigorous models to avoid discriminatory harm.[334]

Similarly, the Institute of Electrical and Electronics Engineers (IEEE), the world's largest organization for technical professionals, released a report entitled Ethically Aligned Design in December of 2016. In that report, the IEEE clearly stated the need for systems to "embed relevant human norms and values."[335]

---

Additionally, the IEEE emphasized the importance of an inclusive approach to stakeholders, relying on tools like explanations or inspection capabilities to increase trust and reliability in machine learning.[336] Here, tools like interactive machine learning or direct questioning and modeling of user responses,[337] "algorithmic guardians" that could help users track and control their shared information,[338] review boards and best practices,[339] multidisciplinary ethics committees,[340] curricula for engineers and technologists that reflects attention to ethical decisionmaking,[341] and the employment of tools like value sensitive or value-based design[342] can go a long way in building a culture of trust, transparency, and accountability in machine learning technologies.

And there are also the commitments made by professional organizations, which are often significant. Of particular note is the code of conduct by the British Computer Society, which holds that individuals must:

> (a) have due regard for public health, privacy, security and wellbeing of others and the environment[;] (b) have due regard for the legitimate rights of Third Parties[;] (c) conduct [their] professional activities without discrimination on the grounds of sex, sexual orientation, marital status, nationality, color, race, ethnic origin, religion, age or disability, or of any other condition or requirement[;] and (d) promote equal access to the benefits of IT and seek to promote the inclusion of all sectors in society whenever opportunities arise.[343]

In turn, perhaps the greatest source of transformation will come from industry's efforts to integrate governance with machine learning models. There is a growing industry developing a number of tools that aim to integrate governance functions into data management systems by focusing on principles like algorithmic transparency (by demonstrating the features used for particular models), using flagging and feedback loops (to address when data or policies change), supporting robust forms of auditing (to study the models being used and their purpose), and developing privacy preserving features (like masking,

---

AUQ9-72N4]. Full disclosure: please note that the author is a member of the IEEE Algorithmic Bias working group.

336. *Id.* at 23.
337. *Id.* at 25.
338. *Id.* at 67.
339. *Id.* at 53.
340. *Id.* at 43.
341. *Id.* at 37–38.
342. *See id.* at 39 (citing SARAH SPIEKERMANN, ETHICAL IT INNOVATION: A VALUE-BASED SYSTEM DESIGN APPROACH (2016)).
343. *Id.* at 43 (citing *BCS Code of Conduct*, BCS, https://www.bcs.org/category/6030 [https://perma.cc/9EFH-WDHB] (asterisks omitted)).

anonymization, and differential privacy).[344] In due time, we could also certainly see entities developing a certification process that draws upon these principles to show their commitment to fairness and transparency, using some of the tools that Josh Kroll and his coauthors have suggested in their work.[345]

## B.    Human Impact Statements in AI and Elsewhere

Algorithmic accountability in private industry also raises a question that underscores the difference between an individualized approach to antidiscrimination and the kinds of issues raised by big data. Title VII approaches take the traditional view, motivated by fairness concerns, that all forms of discrimination are illegal when based on protected categories and must therefore be stamped out.[346] In contrast, big data approaches are less about extinguishing all forms of illegal discrimination; instead, they force us to grapple with the reality that some forms of discriminatory impact may always be present, and focus instead on the question of what efforts can be made to minimize disparate impact.[347]

How can we implement these ideas in the algorithmic context? Recently, a group of prominent researchers launched a document, entitled, "Principles for Accountable Algorithms,"[348] that focused on five core principles: responsibility, explainability, accuracy, auditability, and fairness. They also outlined a host of questions for researchers to explore during the design, prelaunch, and postlaunch phases of algorithmic decisionmaking.[349] Many of their suggested questions focused on various aspects of transparency—for example, identifying parties that are responsible for garnering the social impact of an algorithm, and communicating decisions and describing the sources and attributes of the data

---

344.    All of these tools have been suggested by Matthew Carroll, CEO of Immuta, a company that aims to integrate governance functions into machine learning. *See* Carroll, *supra* note 102.

345.    *See generally* Kroll et al., *supra* note 60 (enumerating the tools, albeit imperfect, for fairness in machine learning as the operationalization of fairness through blindness, statistical parity, fair affirmative action, fair representations, regularization, and fair synthetic data). Indeed, a certification regime that uses these tools well in advance could ensure that fairness is not only a method for virtue signaling, but actually a policy that is both on the books and put into practice.

346.    Barocas & Selbst, *supra* note 60, at 694–95.

347.    *See generally* Andrew D. Selbst, *Disparate Impact in Big Data Policing*, 52 Ga. L. Rev. 109 (2017).

348.    Nicholas Diakopoulos et al., *Principles for Accountable Algorithms and a Social Impact Statement for Algorithms*, FAT/ML, http://www.fatml.org/resources/principles-for-accountable-algorithms [https://perma.cc/6K97-4QB9].

349.    *Id.*

used in machine learning to subjects, including whether it was transformed or cleaned in some manner.[350]

The Principles for Accountable Algorithms also do more than emphasize transparency of authority. They provide important variables to consider in ensuring accuracy and auditability—urging designers to carefully investigate areas of error and uncertainty by undertaking sensitivity analysis, validity checks, and a process of error correction, and also enabling public auditing, if possible, or auditing by a third party, if not possible.[351] Towards this end of encouraging greater collaboration, calibration, consistency, and transparency, we might consider the utility of a "human impact statement," something along the lines of what has been suggested by Marc Roark; a "discrimination impact assessment," as suggested by Selbst; or a "social impact statement," promulgated by a prominent group of algorithmic researchers.[352] Much of the ideas surrounding impact statements originate from environmental law literature,[353] but impact statements have been promulgated in a variety of other areas,[354] also, including human rights,[355] privacy,[356] and data protection.[357] Consider some of the ways in which impact assessments have been used in environmental regulation, which have often served as a transformative blueprint to studying the effect of particular decisions. As described, an environmental impact statement requires the detailed effect of major federal actions on the environment, paying close attention to whether a particular group of people bear a disproportionate share of a negative environmental consequence.[358] Other environmental state statutes, drawing on this principle, also ask for reporting of "any significant physical changes that may be caused by social or economic impacts that are the result of the [p]roject."[359] Environmental impact statements also, like the circumstances here, require in depth research, substantially detailed findings, and can also include a lengthy process of revision, which can last months or even years.[360]

---

350. *Id.*

351. *Id.*

352. *See id.*; Marc L. Roark, *Human Impact Statements*, 54 WASHBURN L.J. 649 (2015); Selbst, *supra* note 347, at 169.

353. *See generally* Selbst, *supra* note 347 (discussing impact statements in policing, drawing on environmental law literature).

354. For excellent commentary on how impact assessments can inform issues that arise from technology and surveillance, see Kenneth A. Bamberger & Deirdre K. Milligan, *Privacy Decisionmaking in Administrative Agencies*, 75 U. CHI. L. REV. 75 (2008); A. Michael Froomkin, *Regulating Mass Surveillance As Privacy Pollution*, 2015 U. ILL. L. REV. 1713; David Wright & Charles D. Raab, *Constructing A Surveillance Impact Assessment*, 28 COMPUTER L. & SECURITY REV. 613 (2012).

355. *See* UNITED NATIONS, HUMAN RIGHTS COUNCIL, OFFICE OF THE HIGH COMM'R, GUIDING PRINCIPLES ON BUSINESS AND HUMAN RIGHTS 23–26 (2011), https://www.ohchr.org/Documents/Publications/GuidingPrinciplesBusinessHR_EN.pdf [https://perma.cc/R3PC-BW5H] (describing human rights impact assessments); *see also* DILLON REISMAN ET AL., AI NOW INST., ALGORITHMIC IMPACT ASSESSMENTS: A PRACTICAL FRAMEWORK FOR PUBLIC AGENCY ACCOUNTABILITY 5 (2018), https://ainowinstitute.org/aiareport2018.pdf [https://perma.cc/JD9Z-5MZC].

356. *See Privacy Impact Assessments*, FED. TRADE COMMISSION, https://www.ftc.gov/site-information/privacy-policy/privacy-impact-assessments [https://perma.cc/C2WF-4PNW] (describing the FTC's system of privacy impact assessments); *see also* REISMAN ET AL., *supra* note 355.

357. *Data Protection Impact Assessments*, ICO, https://ico.org.uk/for-organisations/guide-to-the-general-data-protection-regulation-gdpr/accountability-and-governance/data-protection-impact-assessments [https://perma.cc/Q2NL-9AYZ]; *see also* REISMAN ET AL., *supra* note 355.

At the state level, racial impact statements have been designed to project whether or not a proposed criminal justice legislation will have a disparate racial effect; the general impetus is to discern any racially disparate effects prior to the law's passage or amendment.[361] Typically, a racial impact statement responds to a proposed law that either amends or adds a new crime, by preparing a report that discusses whether the new law will change the state's prison population and/or disproportionately affect minority groups.[362] Unlike environmental impact statements, which require certain actions to be taken in response to an adverse impact, racial impact statements are offered for informational purposes only, even when disproportionate impact is predicted.[363] Since 2007, a number of states have adopted or considered racial impact legislation, some of which is required by the legislature, and others that are initiated by a state sentencing guidelines commission.[364] Other states also adopt a notice and comment period following publication of a racial impact statement.[365] While many states do not

---

require further action after a racially disproportionate finding is made, other states, like Arkansas, Wisconsin, and Kentucky have considered requiring lawmakers to provide an explanation for their course of action, after finding a racial disparity.[366] This has prompted at least one commentator to recommend requiring lawmakers to consider alternative options that may achieve the same policy goals, but without exacerbating racial disparities.[367]

In Europe, the GDPR and Police and Criminal Justice Authorities require data protection impact assessments (DPIA) whenever data processing "is likely to result in a high risk to the rights and freedoms of natural persons."[368] Large scale data processing, automated decisionmaking, processing of data concerning vulnerable subjects, or processing that might involve preventing individuals from exercising a right or using a service or contract, would trigger a DPIA requirement.[369] Importantly, this model extends to both public and private organizations.[370] If a high risk is shown, the organization is required to file a DPIA with the information commissioner's office (ICO) for advice, which the ICO promises to provide within three months.[371]

The DPIA statement is required to reflect four critical areas of attention and is meant to be drafted by the organization's controller, working in conjunction with the organization's Data Protection Officer (DPO).[372] The first is largely descriptive, requiring a description of the processing; the second involves a showing of an assessment of necessity and scale of compliance measures; the third element is identicative, requiring identification and assessment of risks to individuals; and the fourth element is mitigative, requiring a showing of additional measures that could be taken to mitigate risk.[373] Significantly, the controller is in charge of demonstrating GDPR compliance

---

366.  *Id.* at 1464.
367.  *Id.* at 1464 (citing Catherine London, *Racial Impact Statements: A Proactive Approach to Addressing Racial Disparities in Prison Populations*, 29 Law & Ineq. 211, 241 (2011)).
368.  Selbst, *supra* note 347, at 170–71; *see also Data Protection Impact Assessments*, *supra* note 357.
369.  Selbst, *supra* note 347, at 170–71; *see also Data Protection Impact Assessments*, *supra* note 357 (requiring DPIAs if the entity uses "systematic and extensive profiling or automated decision-making to make significant decisions about people," processes data or criminal offence data on a large scale, systematically monitors a publicly accessible place, processes biometric or genetic data, combines or matches data from multiple sources, or processes personal data in a way that involves online or offline tracking of location or behavior, among other categories).
370.  *See* Reisman et al., *supra* note 355, at 7 (making this observation).
371.  *Id.*
372.  *Data Protection Impact Assessments*, *supra* note 357.
373.  *Id.*

and represents a separate entity from the organization that is actually processing the data.[374]

A number of valuable critiques have been raised regarding the execution of DPIAs—for example, they are not required to be published, do not include built-in external researcher review, nor a notice-and-comment proceeding for public review.[375] Yet despite these critiques (most of which are directed towards AI used by public agencies, rather than private corporations), the DPIA process still offers a number of thoughtful insights that can help shape expectations of private companies processing data, just as we see in the GDPR context. The next Subpart discusses some ways to harness the insights derived from impact statements and suggests some elements to consider.

## C.    A Proposed Human Impact Statement

Following the insights offered by other scholars, particularly Selbst and Roark, and the framework offered by the GDPR, as well as other related impact statements,[376] I emphasize three core elements in crafting a Human Impact Statement in Algorithmic Decisionmaking.

First, drawing in part on California's own environmental impact legislation, I recommend the adoption of a substantive, rather than procedural, commitment to both algorithmic accountability and antidiscrimination. In California, the state's Quality Review Act requires "the fair treatment of all races, cultures, and income levels, including minority populations and low-income populations of the state."[377] A statement that assures the public of a commitment to both fairness and accountability, following this example, would go a long way towards setting a baseline set of expectations for private industry to follow.

The second element focuses on the structure of the impact statement and who has responsibility for both implementation and oversight. Here, I recommend the employment of a structure, similar to the GDPR, which relies upon a clear division between the controller (who is responsible for compliance) and the programmer (who is responsible for the algorithm and data processing). By encouraging a healthy separation between the algorithm's designers and

---

those who are tasked to minimize disparate impact, we can ensure greater accountability and oversight.

Third, I also encourage a thorough examination (and structural division), both ex ante and ex post, of both the algorithm and the training data that it is employed to refine the algorithm. As Kroll and his coauthors have observed in an important study, it is possible to demonstrate accountable algorithms through a greater engagement with procedural and technical tools from computer scientists.[378] Ex ante solutions try to correct issues that may surface in the data; ex post solutions try to gather relevant information and reweigh information in order to test the reliability of the data.[379]

Following Andrew Selbst's excellent work on drafting impact assessment for predictive policing techniques, and integrating suggestions derived from literature elsewhere,[380] I emphasize the following specific ex ante criteria:

(1) Identify "potentially impacted populations" and determine their race, ethnicity, gender, sexual orientation, national origin, or other status-based categories;[381]

(2) Identify the effect of uncertainty or statistical error on different groups;[382]

(3) Study whether the decision will have an adverse impact on the subpopulation;[383]

(4) Explore "whether there are reasonable, less discriminatory, alternatives or potential means of mitigation," including the consideration of new target variables or other forms of data, the employment and availability of data processing techniques, and new methods of assessment.[384]

(5) Devote substantial consideration of each alternative in detail so that reviewers can evaluate their comparative merits;[385]

---

378. *See* Kroll et al., *supra* note 60, at 640–41.
379. *See id.* at 637, 662–77.
380. I only mention and summarize these criteria here; Andrew Selbst's discussion, *supra* note 347, is far more detailed and descriptive about the various ways of implementation in an algorithmic context.
381. Roark, *supra* note 352, at 665 n.97 (citing Ramo, *supra* note 359, at 50).
382. Diakopoulos et al., *supra* note 352.
383. Roark, *supra* note 352, at 665 n.94 (citing Ramo, *supra* note 359, at 50); *see also* EPA, Final Guidance for Consideration of Environmental Justice in Clean Air Act 309 Reviews 1 (1999).
384. Roark, *supra* note 352, at 665 n.94 (citing Ramo, *supra* note 359, at 50); Selbst, *supra* note 347, at 173–74.
385. Selbst, *supra* note 347, at 174 (citing 40 C.F.R. § 1502.14(b) (2018)).

(6) Identify and explain the entity's preferred alternative, noting its
    selection among several different algorithmic design choices;[386]

Ex post, an impact assessment should embrace the employment of rigorous techniques and alternatives, to refine and improve the use of AI—its accuracy, its fairness, its accountability, and its transparency. This would include discussion of a set of technical mitigation measures that are not already included in the proposed model.[387] The advantage of employing a rigorous system of impact statements stems from enlisting engineers to explain their design choices, evaluate their efficacy, include alternative configurations, and consider whether any disparate impact has been created for a subpopulation.[388]

Admittedly, these mechanisms may not always be feasible or practical in every instance, but the point of discussing them is to lay the groundwork for reframing the central concern about how big data can impact certain groups, and to create a framework for awareness of these effects. Of course, the cost and length of time it may take to draw up a comprehensive impact assessment may make it difficult to implement.[389] But even aside from cost and complexity, another concern is that without an underlying commitment to a set of normative principles, "impact assessments can become a mere procedural tool that may not be able to create the change they seek," raising the risk that the process may be vulnerable to a host of interests that "may work against the very concerns giving rise to the assessment process itself . . . ."[390] To guard against the possibility of internal self-interest guiding the drafting of an impact statement, we also need to think more broadly about how the law might both incentivize and protect those who come forward.

## V.    REBALANCING TRADE SECRECY THROUGH WHISTLEBLOWING

Part IV dealt with the possibility of industry self-regulation as one potential tool to address algorithmic bias. However, relying on industry self-regulation alone does not address the continued black box problem. As Frank Pasquale noted in The Black Box Society, "[k]nowledge is power. To scrutinize others while avoiding scrutiny oneself is one of the most important forms of power."[391] In countless cases, both inside and outside of the criminal justice system, aggrieved parties have been denied access to the source code that governs them. In the context of big data, Joh details how surveillance technology vendors can block access to even the data that summarizes their results, denoting it to be confidential, proprietary information.[392]  In one public dispute, Palantir Technologies, which had provided the NYPD with software that graphs data derived from the police (arrest records, license plates, parking tickets and the like) in a way that (according to Buzzfeed) "can reveal connections among crimes and people," refused to hand over a readable version of its data to the NYPD after it decided to partner with IBM instead.[393] Even when filing a case of discrimination against a private company remains a possibility, many individuals may not know that an algorithm is discriminating against them, and therefore finding eligible plaintiffs (or crafting a legal theory of illegal conduct) can be difficult, absent some compelling evidence in place.

---

386. *Id.* at 177.
387. *Id.*
388. *See id.* at 173–78.
389. Osagie K. Obasogie, *The Return of Biological Race? Regulating Race and Genetics Through Administrative Agency Race Impact Assessments*, 22 S. CAL. INTERDISC. L.J. 1, 59 (2012).
390. *Id.*
391. PASQUALE, *supra* note 18, at 3 (footnote omitted).

As these observations suggest, a final part of the problem involves trade secrecy.  We continue to view trade secret law as somehow separate from civil rights concerns, and that has contributed to the problem because it has facilitated the absence of accountability.  What is needed instead is a greater recognition of the overlap between these two areas of law.  As David Levine has eloquently explained, on one hand, the very idea of trade secrets invokes both the notion of seclusion intertwined with commerce.[394] At the same time, however, the ideals of democratic government generally aim to minimize commercial interests and the notion of secrecy as a default position.[395]  These tensions—between democratic transparency and commercial seclusion—have become particularly pronounced in the current day, where government has become increasingly intermingled with private industry through privatization and delegation.[396]

The intermingling of public and private, however, is also part of the problem.  It has produced a crisis of transparency, whereby private businesses now play the roles that government used to play, but are able to utilize the principles of trade secret law to protect themselves from the very expectations of

---

transparency that the government operated under.[397] Danielle Citron, nearly ten years ago, observed that the administrative state was slowly being overtaken by closed proprietary systems in areas of public benefits, electronic voting, and agency-gathered data, among others.[398]  Today, the issue is not just that government systems are closed and proprietary—it is also that they are becoming entirely privatized.  David Levine offers several examples—from telecommunications to voting systems—that are now being provided by the private sector, thereby becoming increasingly immunized from transparency by trade secret doctrine.[399]

At the same time, current approaches to regulating algorithms emphasize the need for designers to explain their algorithmic models, rather than disclose them.[400] In 2012, for instance, President Barack Obama proposed the Consumer Privacy Bill of Rights, which would have allowed consumers to challenge and correct data[401] that algorithms use to make decisions about credit or insurance. Congress never acted on it.[402] Both the proposal and any notion that consumers have a right to know what data companies retain about consumers and how that information is used have now disappeared from the White House website.[403]

Some scholars have advocated for greater transparency to expose issues of bias.[404]  Others have taken an alternative route, critiquing transparency as a limited solution that may fail to root out bias.[405] As Joshua Kroll and others have explained, disclosure of source code is only a partial solution to the issue of accountability because of the complexity and dynamism of machine-learning processes.[406]  Some decisions also must necessarily remain opaque to prevent others from gaming the system.[407]  Many systems have also not been designed with oversight and accountability in mind, and thus can be opaque to the outside investigator.[408]    Even auditing has some limitations, depending on the technique.[409]

---

397.  *Id.* at 407–08.

398.  Danielle Keats Citron, *Open Code Governance*, 2008 U. CHI. LEGAL F. 355, 356–57.

399.  Levine, *supra* note 394, at 407.

400.  *See* Bryce Goodman & Seth Flaxman, European Union Regulations on Algorithmic Decision-Making and a "Right to Explanation" (unpublished manuscript) (on file with author).

401.  *See* Press Release, The White House, Office of the Press Sec'y, We Can't Wait: Obama Administration Unveils Blueprint for a "Privacy Bill of Rights" to Protect Consumers Online (Feb. 23, 2012), http://obamawhitehouse.archives.gov/the-press-office/2012/02/23/we-can-t-wait-obama-administration-unveils-blueprint-privacy-bill-rights [https://perma.cc/2TWK-G9JP].

402.  *See* Natasha Singer, *Why a Push for Online Privacy Is Bogged Down in Washington*, N.Y. TIMES (Feb. 28, 2016), http://www.nytimes.com/2016/02/29/technology/obamas-effort-on-consumer-privacy-falls-short-critics-say.html.

403.  The proposal still appears on the Obama White House website.  *See* Press Release, The White House Office of the Press Sec'y, *supra* note 401.

404.  *See* Citron, *supra* note 398, at 358; Schwartz, *supra* note 316, at 1323–25.

405.  Cynthia Dwork & Deirdre K. Mulligan, *It's Not Privacy, and It's Not Fair*, 66 STAN. L. REV. ONLINE 35, 36–37 (2013), https://review.law.stanford.edu/wp-content/uploads/sites/3/2016/08/DworkMullliganSLR.pdf [https://perma.cc/ZCQ5-6N2W].

The cause of this problem, I argue, demonstrates precisely the need for a new approach to trade secrets, in light of the substantial civil rights concerns that algorithms raise.[410]  While I agree with others that accountability is of paramount importance, I would also argue that accountability is impossible without some forms of transparency.  As Anupam Chander has written, "[i]nstead of transparency in the design of the algorithm," we also "need . . . a transparency of inputs and outputs."[411]  In the absence of a centralized, large-scale, federal effort to address this problem, it becomes necessary to explore (1) what solutions might currently exist in the law and (2) whether these solutions might create a platform from which to build further regulatory refinements and encourage greater accountability.  Both of these avenues are only possible, however, with a deeper employment of the limitations of trade secret protection, which are designed precisely to expose potential areas of corporate liability.

The good news, however, is that we have seen variants of this problem before, in other private industries.  As I show in the Subpart below, in other contexts, the law has routinely relied on whistleblowers to address similar information asymmetry and accountability issues.  The same can also be said of algorithmic accountability.  For years, scholars have addressed the need to incentivize internal employees to come forward in cases of significant information asymmetry; those concerns have animated particular provisions in Sarbanes-Oxley, spending statutes, and a host of environmental provisions.  Concerns regarding opacity and difficulty of detection, as I have suggested throughout this Article, are just as salient here in the context of algorithmic accountability, particularly given the biases that can result from skewed data.  As a result, it makes sense to explore the pathways that have been previously taken by legislators, given the potential for similar solutions in this context.

---

406.   Kroll et al., *supra* note 60, at 638–39.
407.   *Id.* at 639.
408.   *Id.* at 649–50.
409.   *Id.* at 650–52.
410.    See Rebecca Wexler's pathbreaking work on this topic, *supra* note 26.
411.    Chander, *supra* note 236, at 1039.

As I show below, a recent, rarely noticed development in modern trade secret law includes federal whistleblowing protections through the employment of the Defend Trade Secrets Act (DTSA) of 2016.[412] I argue that the often overlooked DTSA provisions comprise a hybrid of a solution that could harness the traditional goals and objectives of our language of civil rights, but also immunize whistleblowers to encourage greater transparency in trade secrets.

## A.    The Dominance of Property and the Limits of Transparency

In 1916, beginning with *MacPherson v. Buick Motor Company*,[413] courts began to recognize the importance of extending the notion of accountability and consumer protection to third parties, like family members and bystanders, that were harmed by defective products.[414] According to Jack Balkin, *MacPherson* is particularly on point for the algorithmic age as a case that recognizes the harms that unregulated algorithmic decisionmaking poses, not just to end users, but to other individuals in society as well.[415] Balkin explains how algorithmic models, by externalizing their costs to third parties, cause harm to reputation, produce a lack of transparency in due process, facilitate discrimination, or render individuals more vulnerable to behavioral manipulation.[416] Thus, Balkin argues that algorithm designers should be construed as information fiduciaries because of the dependence between the company that creates the algorithm—the Googles or Facebooks of the world—and the users.[417]

These risks become especially apparent in a world that provides far greater protection to nondisclosure than accountability. Again, property principles pervade systemic disclosure and transparency. Laws such as the Computer Fraud and Abuse Act[418] (CFFA), which has been interpreted to prevent users from violating a website's Terms of Service, have been used to prevent researchers from testing algorithms.[419] Recently, the ACLU sued on behalf of four researchers who maintained that the CFAA actually prevented them from scraping data from sites, or from creating fake profiles to investigate whether algorithmic discrimination led some employment and real estate sites to fail to display certain listings on the basis of race or gender.[420] The concern was that the law permitted researchers to be held criminally accountable because the research might involve violating one of the sites' Terms of Service, something that could carry both prison and fines.[421] As one researcher observed, these laws have the perverse effect of "protecting data-driven commercial systems from even the most basic external analysis."[422]

---

412.    18 U.S.C. § 1836 et seq. (2018). See Peter Menell's groundbreaking work on this topic, *infra* notes 473, 475, and 489.

413.    111 N.E. 1050 (N.Y. 1916).

414.    *See* Jack M. Balkin, *The Three Laws of Robotics in the Age of Big Data*, 78 Ohio St. L.J. 1217, 1232 (2017).

415.    *Id.*

416.    *Id.* at 1238–39.

417.    "Online businesses know a lot about us; we know comparatively little about their operations, and they treat their internal processes as proprietary to avoid theft by competitors." *Id.* at 1228 (footnotes omitted); *see also* Jack M. Balkin, *Information Fiduciaries and the First Amendment*, 49 U.C. Davis L. Rev. 1183 (2016) (discussing the need for a fiduciary relationship between consumers and online platforms).

418.    18 U.S.C. § 1030 (2018).

The researchers had planned to use a variety of different audit testing techniques, including sock puppet profiles and scraping techniques to determine whether certain real estate sites discriminate on the basis of race or other factors.[423] The government, predictably, argued that the case was a purely private matter, characterizing it as a "private actor's abridgement of free expression in a private forum," and questioning the standing of the plaintiffs to file suit.[424] Importantly, the court disagreed with this characterization, noting that "simply placing contractual conditions on accounts that anyone can create . . . does not remove a website from the First Amendment protections of the public Internet."[425]

Since the information it found was already within a public forum (a public website)[426] and was regulated by restrictions on private speech that drew on the imprimatur of state protection through civil or criminal law,[427] it risked state enforcement under the CFAA.[428] Despite this conclusion, which kept the case in court and headed towards trial, the court also found that most of the plaintiffs'

---

activities fell outside of the CFAA, reasoning that "scraping or otherwise recording data from a site that is accessible to the public is merely a particular use of information that the plaintiffs are entitled to see."[429] Although the court reached a different conclusion regarding the creation of fictional user accounts, which would violate the access provision, and therefore raise constitutional considerations, it concluded by reassuring the plaintiffs that the "CFAA prohibits far less than the parties claim (or fear) it does."[430]

At the same time that cases like *Sandvig v. Sessions* provide some optimism for external auditing, it remains necessary to consider the variety of ways in which companies routinely utilize their intellectual property protections to obfuscate inquiry. Even in the context like voting, there have been other cases that, troublingly, demonstrate the power of trade secrets to take precedence over transparency. In 2005, the voting machine company, Diebold Election Systems—now called Premier Election Solutions—refused to follow a North Carolina law that required electronic voting machine manufacturers to place their software and source code in escrow with a state Board of Elections approved agent.[431] Over a series of court battles, Diebold refused to comply, eventually withdrawing from the state altogether, rather than reveal its source code.[432] In another event, also discussed by Levine, when hackers successfully accessed (and manipulated) a series of Diebold machines, Diebold chose to characterize the events as "potential violations of licensing agreements and intellectual property rights," rather than responding to it as a threat to the dignity of the voting tabulation process.[433]

The risks become especially evident in an era where corporations have become especially dependent on trade secret protection where algorithms are concerned. Because the code for a machine learning algorithm is so complex, simply reading it does not make it interpretable without the ability of interpreters to plug in data and see how the model actually functions.[434] Further, because algorithmic models often depend on the input of unique personal data, the outcomes may be obscure and difficult to study in a collective capacity.[435] Bias, error, and faulty assumptions plague the design of algorithms as a result of humans designing those algorithms. Few could spot errors in code by reading a description of how that code ought to function. Similarly, few defendants can explain why an algorithmic model predicted recidivism for them without an opportunity to examine why it reached such predictions. Only other humans who understand the programming languages and statistical models that underlie algorithms can pinpoint those errors by examining them.

---

429. *Id.* at 26–27. It found that employing a bot to crawl through web sites might violate a website's Terms of Service, but it did not constitute an "access" violation per se "when the human who creates the bot is otherwise allowed to read and interact with that site." *Id.* at 27.

430. *Id.* at 34.

431. Levine, *supra* note 394, at 419–20. For an excellent article exploring the use of software-independent voting systems, compliance audits, and risk-limiting audits in elections, see Philip B. Stark & David A. Wagner, *Evidence-Based Elections*, 10 IEEE SECURITY & PRIVACY 33 (2012).

432. Levine, *supra* note 394, at 420.

433. *Id.* at 421 (quoting Ion Sancho, the Supervisor of Elections in Leon County, Florida, where the hacks took place). "I really think they're not engaged in this discussion of how to make elections safer." *Id.* (footnote omitted).

Software companies, however, currently have other ways to protect their intellectual property and guard the value of their products. Software patents once encouraged companies like Northpointe to disclose algorithms to exclude direct competitors.[436] After a golden age of trolls and overenforcement in the early twentieth century, however, Supreme Court decisions—such as *Bilski v. Kappos* and *Alice Corp. v. CLS Bank International*—have essentially ended patent protection for software like COMPAS.[437] Disclosing a way of assessing recidivism with a computer to the United States Patent and Trademark Office would unlikely be worth Northpointe's time and trouble, given the dubious protection that software patents now receive.

Copyright laws create a similar problem. Complex algorithms essentially boil down to a string of commands. Copyright laws protect these strings of commands, as they would any other string of syntax, as a literary work. Consequently, only its precise expression, the names of commands, for instance, is protected.[438]

---

434. Christian Sandvig et al., *Auditing Algorithms: Research Methods for Detecting Discrimination on Internet Platforms* 10 (May 22, 2014), http://www-personal.umich.edu/~csandvig/research/Auditing%20Algorithms%20--%20Sandvig%20--%20ICA%202014%20Data%20and%20Discrimination%20Preconference.pdf [https://perma.cc/V8ED-R83M] (presented at 64th Annual Meeting of the International Communication Association).

435. *Id.*

436. *See* Gene Quinn, *Building Better Software Patent Applications: Embracing Means-Plus-Function Disclosure Requirements in the Algorithm Cases*, IPWATCHDOG (June 18, 2012), http://www.ipwatchdog.com/2012/06/18/building-better-software-patent-applications-embracing-means-plus-function-disclosure-requirements-in-the-algorithm-cases/id=24273 [https://perma.cc/JAC3-FKB8].

437. *See* Alice Corp. v. CLS Bank Int'l, 134 S. Ct. 2347, 2349–51 (2014); Bilski v. Kappos, 561 U.S. 593, 593–96 (2010) (limiting the scope of patentability over software-related inventions).

438. It is somewhat ironic that criminal courts give so much more protection to software code secrecy than it would receive in a civil case. Interestingly, Northpointe does not even guarantee that COMPAS does not infringe other's intellectual property rights, even while In part because of the shortcomings of copyright and patent protection, trade secrets have become the default way to protect algorithms and the source code that embodies them. Although trade secret law remains perhaps the only reasonable way to protect source code, it is also a poor way to protect the public interest. To be a trade secret, information must: (1) not be generally known, (2) bring economic value to its holder by virtue of not being publicly known, and (3) be subject to reasonable precautions to keep the information secret.[439] If information is already known or is even readily ascertainable, it cannot be a trade secret.[440] Federal statutes involve the Economic Espionage Act, which instituted the first federal scheme for trade secret protection and also introduced criminal penalties for misappropriation.[441]

Yet trade secrets, particularly in the software context, suffer from a paradox. As some have observed, without first disclosing and examining the source code, it is impossible to know whether an algorithm even qualifies as a trade secret.[442] But disclosure would potentially jeopardize its status as a trade secret. To avoid this issue, most entities simply assert trade secrecy even when the underlying information may not actually qualify as a trade secret. There is no way to tell otherwise, absent some form of disclosure. Largely because of the deference that companies enjoy in this context, information-based products have long favored trade secret protection, which has led to some scholarly debates and discussion.[443]

---

maintaining that secrecy is essential to its business. *See* COMPAS Licensing Agreement § 8.2 (2010), https://epic.org/algorithmic-transparency/crim-justice/EPIC-16-06-23-WI-FOIA-201600805-2010InitialContract.pdf [https://perma.cc/6Y6H-XN9M]; Katyal, *supra* note 25.

439. *See* 18 U.S.C. § 1839(3)(B) (2018); Metallurgical Indus., Inc. v. Fourtek, Inc., 790 F.2d 1195, 1199 (5th Cir. 1986).

440. *See* Uniform Trade Secrets Act § 1(4) (Nat'l Conference of Comm'rs of Unif. State Laws 1985).

441. Economic Espionage Act of 1996, Pub. L. No. 104-294, 110 Stat. 3488 (codified as amended at 18 U.S.C. §§ 1831–1839 (2018)).

442. *See* Charles Short, *Guilt by Machine: The Problem of Source Code Discovery in Florida DUI Prosecutions*, 61 Fla. L. Rev. 177, 190 (2009) (discussing State v. Chun, 923 A.2d 226 (N.J. 2007), where the code underlying supposedly proprietary breathalyzer software was revealed to consist primarily of general algorithms that arguably would not qualify as a trade secret).

443. *See* Michael Mattioli, *Disclosing Big Data*, 99 Minn. L. Rev. 535, 550 (2014) (citing Mark A. Lemley & David W. O'Brien, *Encouraging Software Reuse*, 49 Stan. L. Rev. 255, 258 (1997) (noting the use of trade secret protection in software industry); Peter S. Menell, *The Challenges of Reforming Intellectual Property Protection for Computer Software*, 94 Colum. L. Rev. 2644, 2652 (1994) (same)). Initially, some scholars argued that by keeping their information secret, companies were slowing the pace of innovation by engaging in potentially duplicative projects at one time. *Id.* at 551 (citing Robert G. Bone, *A New Look at Trade Secret Law: Doctrine in Search of Justification*, 86 Calif. L. Rev. 241, 266–67 (1998)).

## B.    Whistleblowing and Secrecy

In a powerful white paper from the Future of Humanity Institute, authors Miles Brundage and Shahar Avin wrote about the need to promote a culture of responsibility in AI.[444] One of their suggestions for future research involved the enlistment of whistleblowing protections, pointing out the need to explore its potential intersection with preventing AI-related misuse.[445]

Whistleblowing activity involves "the disclosure by an organization member '(former or current) of illegal, immoral or illegitimate practices under the control of their employers, to persons or organizations who may be able to effect action.'"[446] The whistleblower, in this case, might be someone from within who is motivated by a concern for others' wellbeing, and who can shed light on the algorithm, its projected or actual impact, and also importantly, the data that it is trained upon, to determine whether bias is an issue. This Subpart outlines how as a general matter, whistleblower protections might affect the context of algorithmic accountability, by protecting individuals who may come forward to address issues of discrimination and bias. Of course, this protection is only a partial solution, given the opacity of trade secrecy, but as I argue below, it does provide some form of protections for those who choose to come forward.

Whistleblowing protections have been employed in a wide variety of models that range from incorporating antiretaliatory whistleblower protections into regulatory statutes, to private rights of action for whistleblowers, to offering monetary incentives to those who report wrongdoing.[447] In 1989, Congress unanimously passed the Whistleblower Protection Act (WPA) and amended it five years later.[448] The Act discourages "employer retaliation against employees

---

Others, like Mark Lemley, postulated that the legal protection for trade secrets in software would mean less investment in physical barriers to access (like encryption), and perhaps would encourage greater information sharing as a result. *Id.* at 552 (citing Mark A. Lemley, *The Surprising Virtues of Treating Trade Secrets as IP Rights*, 61 STAN. L. REV. 311, 333–34 (2008)).

444.  MILES BRUNDAGE & SHAHAR AVIN, THE MALICIOUS USE OF ARTIFICIAL INTELLIGENCE: FORECASTING, PREVENTION, AND MITIGATION 56 (2018), https://img1.wsimg.com/blobby/go/3d82daa4-97fe-4096-9c6b-376b92c619de/downloads/1c6q2kc4v_50335.pdf [https://perma.cc/49CN-DCTM].

445.  *Id.*

446.  Peter B. Jubb, *Whistleblowing: A Restrictive Definition and Interpretation*, 21 J. BUS. ETHICS 77, 84 (1999) (citing MARCIA P. MICELI & JANET P. NEAR, BLOWING THE WHISTLE 15 (1992)).

447.  *See* Orly Lobel, *Citizenship, Organizational Citizenship, and the Laws of Overlapping Obligations*, 97 CALIF. L. REV. 433, 442–43 (2009).

448.  *See* Stephen R. Wilson, *Public Disclosure Policies: Can a Company Still Protect Its Trade Secrets?*, 38 NEW ENG. L. REV. 265, 270 (2004).

who report violations concerning fraud, waste, or abuse."[449]  Although the WPA was motivated by a desire to create a protected class of government employees, the government has since included whistleblower protections in over fifty other federal statutes, extending to private entities in its purview.[450]  The Sarbanes-Oxley Act, for example, protects employees who reveal evidence of corporate fraud to an appropriate state or federal authority.[451]  Most statutes impose strong penalties on employers who retaliate by discharging or discriminating against the whistleblowing employee by awarding them reinstatement, along with substantial amounts of damages, among other awards.[452]

There are three potential arguments for paying greater attention to whistleblowing in this context.  The first, and most important, considers the barrier of intellectual property protections, which often secludes crucial information.  Given the issues of opacity, inscrutability, and the potential role of both trade secrecy and copyright law in serving as obstacles to disclosure, whistleblowing might be an appropriate avenue to consider in AI.[453]  Whistleblowing has also been shown to be particularly effective in similar situations that involve information asymmetry (for example in cases of corporate wrongdoing), where whistleblowers have been considered to be vital to achieving greater compliance because they can help to detect areas of wrongdoing.[454]  Research suggests that individuals (like Christopher Wylie discussed in the Introduction) may be motivated by a belief that whistleblowing constitutes "conscience cleansing," to advance social welfare, or because they are discontented in some way.[455]

---

449.  *Id.*

450.  *Id.* at 271.

451.  *Id.* at 272.

452.  *Id.* at 271–72.  Some statutes, which are known as core statutes, focus primarily on the protection of whistleblowing activities, and others are adjunct statutes because they protect whistleblowing activities within the context of another primary, legislative purpose.  Some statutes provide no more than a cause of action to whistleblowers who experience retaliation for their activities; others provide a financial reward, in addition to employee protection. Bishara et al., *infra* note 455, at 44 (comparing the New Jersey Conscientious Employee Protection Act and the Clean Air Act).

453.  *See* Levendowski, *supra* note 56 (discussing how copyright protection for data can serve as an obstacle to improving data quality).

454.  *See, e.g.*, Stuart Lieberman, *Whistleblowers Can Prevent Toxic Nightmares*, LIEBERMAN & BLECHER, https://www.liebermanblecher.com/aop/slapp-suit-and-environmental-whistblower/ environmental-whistleblower-cases [https://perma.cc/8MMM-G3JD] (cited in Anthony Heyes & Sandeep Kapur, *An Economic Model of Whistle-Blower Policy*, 25 J.L. ECON. & ORG., 157, 161 (2008)) (observing that "Whistleblower laws are particularly helpful in environmental cases.  This is so because many environmental violations and crimes are difficult to detect absent help from knowledgeable insiders").

Second, whistleblowing protections are particularly appropriate, where, as here, the government is relying more and more on private entities for its various governing activities. As I discussed in a related paper, as privatization and delegation becomes the norm for our world of automated governance, it becomes even more necessary to explore ways to incentivize individuals to come forward.[456] Orly Lobel, in her extensive work on whistleblowing, has argued that "as government relies more on private parties to prevent improper behavior, the need for legal protections for whistleblowers increases."[457] Those conditions are especially appropriate here, where, as discussed earlier, automated decisionmaking systems have become arbiters for a host of government decisions, including Medicaid, child-support payments, airline travel, voter registration, and small business contracts.[458]

Finally, and perhaps most important, whistleblowing has been shown to be particularly effective in situations, like this one, where companies are increasingly relying on internal systems of self-regulation and trying to address the importance of combating bias. Here, particularly given the internal nature of AI, there is even more of a necessity to integrate a culture of whistleblower protection. And it is important to observe, as the Wylie example shows at the start of this Article, that people can be motivated to come forward, even in the absence of monetary reward.[459] In a similar context, Orly Lobel has found that

---

455. Heyes & Kapur, *supra* note 454, at 164–71. Other variables that influence whistleblowing include the following: (1) confidence that their organization would address the wrongdoing, (2) the belief that the organization supported whistleblowing, in general, (3) the seriousness of the allegation, (4) the whistleblower's desire to "put 'their' organization on the right track," and (5) the availability of a monetary reward. Norman D. Bishara et al., *The Mouth of Truth*, 10 N.Y.U. J.L. & Bus., 37, 60 (2013).

456. *See* Katyal, *supra* note 25.

457. Lobel, *supra* note 447, at 473.

458. *See generally* Citron, *supra* note 398.

459. In a powerful experiment, Lobel and Feldman used a series of experimental surveys among 2000 employees, and asked them to predict their own actions (as well as the actions of others) when confronted with an illegal scheme whereby a company defrauded the government by overcharging, and then undersupplying regarding a construction contract, causing some risk to the public and reducing government funds  The authors then studied the values that employees assigned to different regulatory mechanisms and the legal incentives assigned to prompt them to come forward.  The authors' findings led them to conclude that "when noncompliance is likely to trigger strong internal ethical motivation, offering monetary rewards may be unnecessary, or, worse yet, counterproductive." She also points out that in situations where an unlawful act is perceived to be morally offensive, a duty to report may be all that is needed to encourage folks to come forward.  However, if there is no internal motivation present (like if the misconduct seems low in severity), then external incentives, like material incentives, were more influential in incentivizing people to whistleblower protections are often necessary as complements to systematic self-monitoring.[460] Lobel concludes that "for certain types of misconduct, policymakers should consider ways to instill ethical norms about different regulatory fields," using educational programs and improving communication channels as part of this project.[461] Ultimately, she opts for a model that prioritizes internal reporting over external reporting, but notes that if the internal process is nonresponsive, then "it becomes reasonable for an employee to step outside the organization."[462]

Whistleblower advocates have argued that whistleblowing activities actually conserve law enforcement resources, because it increases both the speed of detection and correction, far more than an external source of monitoring, and promotes internal self-monitoring to the extent that organizations are aware of the possibility of exposure.[463] Whistleblowing can be an important, efficient, and valuable source of feedback, particularly in cases of intra-organizational disclosures, because it can correct misunderstandings and wrongdoing without the financial and reputational risks associated with public disclosure.[464]

## C. Trading Secrecy for Transparency

In 2016, the Federal government, recognizing the confusion and uncertainty that characterized state trade secret laws, as well as the importance of trade secrets to an economy dependent on information,[465] passed the Defend

---

come forward. *See* Orly Lobel, *Linking Prevention, Detection and Whistleblowing: Principles for Designing Effective Reporting Systems*, 54 S. Tex. L. Rev. 37, 46–47 (2012) (detailing study in Yuval Feldman & Orly Lobel, *The Incentives Matrix*, 88 Tex. L. Rev. 1151, 1176 (2010)).

460. Orly Lobel, *Lawyering Loyalties: Speech Rights and Duties Within Twenty-First Century New Governance*, 77 Fordham L. Rev. 1245, 1249 (2009). A system that relies primarily on external reporting, Lobel argues, has its own set of limitations, stemming mostly from the reality that most individuals are reluctant to report misconduct to an outside agency, particularly given the material and social risks of disclosure. Lobel, *supra* note 459, at 43.

461. Lobel, *supra* note 459, at 47.

462. Lobel, *supra* note 447, at 492. *See also* Lobel, *supra* note 460 (examining the role of the lawyer in whistleblowing).

463. Bishara et al., *supra* note 455, at 39–40.

464. *Id.* at 40.

465. In a press release announcing the 2014 version of the Defend Trade Secrets Act (DTSA) Senator Hatch warned:

> In today's electronic age, trade secrets can be stolen with a few keystrokes, and increasingly, they are stolen at the direction of a foreign government or for the benefit of a foreign competitor. These losses put U.S. jobs at risk and threaten incentives for continued investment in research and development. Current federal criminal law is insufficient.

Trade Secrets Act[466] (DTSA) in early 2016 with little serious opposition.[467] It amended the Economic Espionage Act (EEA) to create a private cause of action for the EEA's trade secret provisions.[468] The DTSA also authorized enforcement of violations of state trade secret protections "related to a product or service used in, or intended for use in, interstate or foreign commerce" in federal court.[469] Under the DTSA, federal courts can grant ex parte orders for preservation of evidence and seizure of any property used to commit or facilitate a violation of the statute, a remedy much more powerful than what was previously available under state trade secret laws.[470]

Importantly, Congress also recognized that strong trade secret protection can threaten the public interest.[471] Consequently, the DTSA also immunizes whistleblowers from liability under federal and state trade secret law for disclosure, in confidence, of trade secrets to government officials and attorneys for the purpose of reporting a possible violation of law.[472] At the heart of recent federal law protecting trade secrets, for example, lies an allowance that provides for immunity from trade secret liability for a confidential disclosure to government officials and attorneys for the purpose of reporting or investigating a suspected violation of law.[473] The DTSA whistleblower immunity regime aims to hold companies accountable for possible misconduct by allowing authorities

---

to scrutinize trade secrets without damaging legitimate trade secret owners.[474]
The provision was allegedly designed, in part, to follow Peter Menell's
pathbreaking work tying trade secrecy to the need for a public policy exception
to protect whistleblowing activity, published in the California Law Review in
2017.[475]

Immunity was needed, the government realized, in order to encourage
greater accountability, since the threat of liability for trade secret
misappropriation might deter individuals from coming forward.[476]   Senate
Judiciary Committee Chairman Charles Grassley stated:

> Too often, individuals who come forward to report wrongdoing in the
> workplace are punished for simply telling the truth.    The
> amendment . . . ensures that these whistleblowers won't be slapped
> with allegations of trade secret theft when responsibly exposing
> misconduct.   It's another way we can prevent retaliation and even
> encourage people to speak out when they witness violations of the
> law.[477]

And there is evidence to suggest that employees may be the best source of this
information.  In the context of fraud, for example, nearly 40 percent of cases of
initial fraud detection came from employee tips, as compared to 16.5 percent
from internal audits and 13.4 percent from management review.[478]

Given these statistics, it is possible to imagine a world where internal
employees, after considering the impact of an algorithmic model on particular
groups, might feel protected under the DTSA to come forward to address issues
that could give rise to antidiscrimination or privacy concerns.  At the very least,
they may reach out to lawyers and others to determine whether a violation may
have occurred.  In the context of algorithmic bias, a whistleblower can be crucial
to shedding light on the potential implications of an algorithm on social groups,
particularly minorities, and also on other entitlements, like informational
privacy.

One could imagine an employee at any major tech company, for example,
noting the potential for disparate impact in algorithmic decisionmaking and
making attempts to ensure some form of legal accountability as a result of their

---

discovery. And if this seems like a far-fetched idea, it is well worth remembering how successful whistleblowing has been in other, comparable contexts of corporate wrongdoing, in the fraud and environmental arenas.[479]  There is no reason to believe that similar exemptions will not have at least some positive effect in encouraging accountability in the algorithmic context as well.

Although the DTSA strengthens the remedies for trade secret misappropriation, it also balances this approach by granting immunity to would-be informants who reveal confidential information to attorneys or government investigators for the purposes of investigating violations of law.[480] Companies are actually required to notify employees if they comply with the DTSA criteria, as employees are entitled to immunity in any contract that governs the use of trade secrets.  As one source further explains:

> Specifically, Section 7 of the DTSA provides criminal and civil liability to any person who discloses a trade secret under two discrete circumstances: (1) when the disclosure is made in confidence to a government official or attorney for the sole purpose of reporting or investigating a suspected violation of the law, and (2) when the disclosure is made in a complaint or other document filed under seal in a judicial proceeding.[481]

Here, the DTSA fits in with other whistleblowing protections: (1) the False Claims Act, which was enacted to deter fraud against the government, (2) the Sarbanes-Oxley Act, which instituted its own whistleblowing protections to encourage others to come forward in cases of corporate wrongdoing, and (3) the Dodd-Frank Wall Street Reform and Consumer Protection Act, which encourages the reporting of securities violations, among other elements.[482]

Yet what sets the DTSA provisions apart is significant.  First, the DTSA does not require—nor even envision—public disclosure of the trade secret.  The veil of partial secrecy supports the idea, advanced by Ed Felten, that "[t]ransparency needn't be an all-or-nothing choice."[483]  As some have argued, the prospect of regulatory transparency, even as a general matter, can be tremendously costly

---

from an administrative perspective.[484]   For one thing, the sheer complexity, magnitude, and dynamism of algorithms and machine learning practices make comprehension of the code—and the categories that shaped it—incredibly difficult.[485]  However, under a DTSA procedure, the trade secret is under seal and largely secure from public view at all times.  The only individuals charged with the ability to view the trade secret are the government, the individual whistleblower, and the whistleblower's attorney.[486]

Second, the DTSA envisions a carefully calibrated approach where a whistleblower must come forward in order to instigate an investigation, ensuring that other responsible parties (an attorney or government official) can then also play a role in investigating whether a violation occurred.[487]  In other words, the statute grants immunity to those persons who theoretically are most likely to understand the effects of the algorithm, thereby reducing the information asymmetry that outside researchers may face.[488]  The advantage of this process ostensibly ensures that potential allegations are carefully explored before any legal action is taken, and that algorithms are always behind the protected veil of secrecy or under seal in court.  This case-by-case approach suggests a greater level of specificity, given that an employee would only come forward if she had a reasonable prospect of believing that a legal violation had taken place, since the statute does not provide them with any benefits other than immunity.  The advantage of the DTSA procedure is that it augments, but does not replace, the discussions about the need for broad regulatory transparency in an algorithmic age by providing a case-by-case opportunity for clarity and accountability.

Third, the virtue of the DTSA whistleblower immunity lies in its employment of trusted intermediaries—government officials bound by state and federal law to protect trade secrets and attorneys bound by ethical obligations of confidentiality—to protect against the risk of commercial harm to legitimate trade secret owners.[489]   The government has a long tradition of requiring disclosure of data that is protected by trade secrets when it raises important public policy concerns. The federal government has effective safeguards in place for protecting the confidentiality of trade secret information.[490] Patent applications are kept in confidence by the Patent and Trademark Office; the Food and Drug Administration reviews drug applications, keeping clinical trial data and manufacturing methods in secret; the Securities and Exchange Commission protects confidential business information; and even the Freedom of Information Act, the regulation most committed to open government, steadfastly exempts trade secrets from public disclosure.[491] Should the government violate trade secret protection, courts would allow individual owners to bring takings lawsuits against the government under the Fifth Amendment.[492] Thus, there is no reason why we would not apply the same standards in an age of algorithmic accountability.

---

484. *See* Maayan Perel & Niva Elkin-Koren, *Black Box Tinkering: Beyond Disclosure in Algorithmic Enforcement*, 69 FLA. L. REV. 181, 187 (2017).

485. *See id.* at 188.

486. Of course, the hiring of experts to audit the trade secret to determine possible liability may require additional considerations not yet envisioned by the statute, but some arguments for the extensions of immunity may at least be arguably warranted here.

487. *See* 18 U.S.C. § 1833 (2018).

488. For general economic theory around assigning whistleblower immunity, see Heyes & Kapur, *supra* note 454.

489. *See* Menell, *supra* note 475, at 56, 60; *see also* Menell, *supra* note 473; Peter S. Menell, *The Defend Trade Secrets Act Whistleblower Immunity Provision: A Legislative History*, 1 BUS. ENTREPRENEURSHIP & TAX L. REV. 398 (2018).

Fourth, unlike the specific provisions of the whistleblowing provisions in the Fair Credit Act or Sarbanes-Oxley, which are calibrated to specific kinds of legal wrongdoing, the DTSA's main virtue lies in its broad reference to "violation of law,"[493] however broadly defined. This means, at least conceivably, that violations of the Federal Trade Commission (FTC) Act, which protects against unfair or deceptive business practices,[494] would arguably fall within its purview. In the past, the FTC has used its authority to respond to behavioral marketing concerns, to regulate the rising authority of influencers, and to develop a set of best practices for private sector cybersecurity.[495] But even aside from the FTC's broad statute, a host of other statutes—the Health Insurance Portability and Accountability Act; the Children's Online Privacy Protection Act of 1990; the Fair and Accurate Credit Transactions Act of 2003; and the Family Educational Rights and Privacy Act—all implicate informational privacy.[496]

A fifth consideration is worth discussing, particularly in areas beyond informational privacy. Just as in the context of government fraud under the FCA, the very presence of the DTSA provisions can encourage companies to be more accountable, particularly for the purposes of avoiding the triggering of a whistleblowing event.[497]   This may be the case even when the precise legal violation may not be clear.  For example, while the limitations surrounding Title VII have been eloquently explored by Selbst and Barocas, the risk of a DTSA whistleblowing event might still encourage companies to remain vigilant against discriminatory treatment, for the purposes of assuring both their employees and the public of their commitment to nondiscrimination.[498]

In the context of algorithms, we might see how the role of a whistleblower can contribute to the goal of nondiscrimination.  As a general matter, the DTSA requirement that every transaction is required to provide notice to every relevant employee can arguably reflect a broader, more cultural attentiveness to compliance with existing law.   Moreover, the presence of a potential whistleblower creates the prospect of both direct and indirect surveillance over the internal activities of algorithmic design.  Even if the sanctions are unclear, diffuse, or uncertain, the prospect of a whistleblower might create the incentives to respond to prospective disparate impacts.

Recent cases illustrate the risks involved in trusting those who write algorithms with widespread effects to ensure their efficacy.  In Italy, for example, a programmer who wrote the software that timed traffic lights may have conspired with government officials, police officers, and seven private companies to rig the traffic lights.[499]   The lights would stay yellow for an unusually brief period, thus catching more motorists in the red.[500]   The deceit came to light only after the unusually high number of red light tickets drew official scrutiny.[501]   In 2015, it was revealed that Volkswagen programmed its diesel vehicles to perform differently during emissions tests by regulators than on the road.[502]   Because the software was proprietary, however, it was shielded from outside scrutiny. Under the cloak of trade secrets, Volkswagen used its source code to potentially defraud consumers and regulators for years.[503]

---

497. *See, e.g.*, *How to Avoid False Claims Act Liability—What Every Compliance Officer Needs to Know*, GIBSON DUNN (Mar. 26, 2010), http://www.gibsondunn.com/publications/Pages/HowtoAvoidFalseClaimsActLiability.aspx [https://perma.cc/ERF2-33JZ] (instructing clients how to avoid FCA liability).

498. *See* Barocas & Selbst, *supra* note 60.

499. Jacqui Cheng, *Italian Red-Light Cameras Rigged With Shorter Yellow Lights*, ARS TECHNICA (Feb. 2, 2009, 6:15 PM), https://arstechnica.com/tech-policy/2009/02/italian-red-light-cameras-rigged-with-shorter-yellow-lights [https://perma.cc/VM3K-EH38].

500. *Id.*

501. Sergey Bratus, Ashlyn Lembree & Anna Shubina, *Software on the Witness Stand: What Should It Take for Us to Trust It?*, *in* TRUST AND TRUSTWORTHY COMPUTING 396, 404 (Alessandro Acquisti et al. eds., 2010) ("[H]ad the bias been less pronounced, it might have not been detected at all.").

502. David Kravets, *VW Says Rogue Engineers, Not Executives, Responsible for Emissions Scandal*, ARS TECHNICA (Oct. 8, 2015, 10:40 AM), http://arstechnica.com/techpolicy/2015/10/volkswagen-pulls-2016-diesel-lineup-from-us-market [https://perma.cc/62DU-8V4W].

Although not as dramatic or craven, errors have already appeared in the algorithms used in the criminal justice system. In New Jersey, a court ordered a software developer to disclose the source code for a breathalyzer.[504] "[C]atastrophic error detection [was] disabled" in the software so it "could appear to run correctly while actually executing invalid code."[505] In 2016, a New York state court refused to admit evidence analyzed by the STRmix algorithm due to issues raised with its accuracy.[506] The error reduced the probability that a DNA sample matched a given defendant in certain circumstances.[507] Because the mistake was in a conditional command, it happened rarely and was thus difficult for even the developer to detect.[508] Had the algorithm's source code remained a secret, the error would have never been discovered.

These examples illustrate the significant potential for the use of existing law as a public policy carveout to provide for the limited disclosure of trade secrets in situations of potential algorithmic bias. Companies who write proprietary software must likewise be accountable when their algorithms produce disparate treatment in decisionmaking, particularly given the risk that their employees may be revealing these issues to third parties under the DTSA. Under the DTSA, individuals can and should feel empowered to turn over source code to an attorney or federal employee—both of whom are bound by confidentiality obligations—so that they can fully examine the algorithm's operations, accompanying logic, and protect the interests of the public. An allowance for whistleblower protection might serve either as a pathway to address algorithmic accountability at the federal level or as an incentive to encourage companies themselves to remain vigilant against the prospect of making illegal decisions based on race or other protected characteristics.  In either case, it is a valuable tool to partially address the problem.

---

503. *See* Jake J. Smith, *What Volkswagen's Emissions Scandal Can Teach Us About Why Companies Cheat*, KELLOGGINSIGHT (Feb. 2, 2017), https://insight.kellogg.northwestern.edu/article/what-volkswagens-emissions-scandal-can-teach-us-about-why-companies-cheat [https://perma.cc/G74L-9EHP].

504. Ryan Paul, *Buggy Breathalyzer Code Reflects Importance of Source Review*, ARS TECHNICA (May 15, 2009, 7:57 AM), https://arstechnica.com/tech-policy/2009/05/buggy-breathalyzer-code-reflects-importance-of-source-review [https://perma.cc/GU3T-G4CS].

505. Short, *supra* note 442, at 185 (footnote omitted).

506. *Id. See* Jesse McKinley, *Judge Rejects DNA Test in Trial Over Garrett Phillips's Murder*, N.Y. TIMES (Aug. 26, 2016), https://www.nytimes.com/2016/08/27/nyregion/judge-rejects-dna-test-in-trial-over-garrett-phillipss-murder.html; *Ruling—the People of the State of New York versus Oral Nicholas Hillary (NY): DNA Evidence Admissibility*, STRMIX (Sept. 12, 2017), https://strmix.esr.cri.nz/news/ruling-the-people-of-the-state-of-new-york-versus-oral-nicholas-hillary-ny-dna-evidence-admissibility [https://perma.cc/2DPA-7ZUF] [hereinafter *Ruling—the People of the State of New York*].

507. *See Ruling—the People of the State of New York*, *supra* note 506.

508. *Id.*

## D.    Some Significant Caveats

There are, of course, some very important qualifications to draw here.  The first, and most important, involves a central challenge to transparency itself.  As many scholars have noted, source code disclosure is just a partial solution to the problem of algorithmic accountability.[509]  It is hard to know, as a general matter, whether something is potentially unlawful, particularly given the grey areas of legal interpretation.[510]  A limited disclosure of an algorithm tells you very little, because its effects cannot be interpreted by a simple reading of the code.[511]  As Christian Sandvig explains:

> Algorithms also increasingly depend on personal data as inputs, to a degree that the same programmatically-generated Web page may never be generated twice.  If an algorithm implements the equation resulting from a multivariate regression, for instance, with a large number of variables it becomes virtually impossible to predict what an algorithm will do absent plugging in specific values for each variable. This implies that some badly-behaving algorithms may produce their bad behavior only in the context of a particular dataset or application, and that harmful discrimination itself could be conceptualized as a combination of an algorithm and its data, not as just the algorithm alone.[512]

To compensate for this problem, investigators have to plug data into the algorithm in order to see how it operates.[513]

Even aside from the general issue regarding interpretation, there are other objections to draw.  Perhaps the most obvious is that a whistleblower provision only partially addresses the problem of algorithmic accountability.  It is, admittedly, an imperfect first step towards pulling back the veil of trade secrecy over source code.  And, as Kroll and his coauthors note, full transparency is not always possible, or even desirable, if it amounts to destruction of a trade secret or revelation of sensitive or protected data.[514] At other times, it can lead to other undesirable effects, like gaming of the system.[515] And the problem may not always be secrecy or opacity; as Selbst and Barocas have remind us, the problem may actually reside in systems that are inscrutable, because they make it impossible for a human to intuitively reason about how they operate.[516] As Kate Crawford and Mike Annany further elaborate, the very notion of transparency suggests that one can glean insights from observing a set of results, and then hold systems accountable for their choices.[517] But there are different types of opacity at work here.[518] As Jenna Burrell reminds, one kind involves the notion of intentional concealment; another involves the complexity and specialization of the information; and another involves the complexity of machine learning itself.[519]

---

509.  Kroll et al., *supra* note 60, at 638–39.
510.  *See* Lobel, *supra* note 447, at 464 ("Employees frequently face possible illegal behavior, but the degree of unlawfulness is usually open to interpretation.").
511.  Sandvig et al., *supra* note 434, at 10.
512.  *Id.*
513.  *See id.*; *see also* Deven R. Desai & Joshua A. Kroll, *Trust But Verify: A Guide to Algorithms and the Law*, 31 Harv. J.L. & Tech. 1, 10 (2017) (noting that auditing "can only test 'a small subset of potential inputs'" (quoting Kroll et al., *supra* note 60, at 650)).

Another cluster of objections to the DTSA, demonstrated by some other areas of case law, is that a broad public policy in favor of whistleblowing activities might justify aggrieved employees to engage in a "fishing expedition" prior to their discharge that might lead them to carry off proprietary information, including the data that an algorithm was trained upon. In one such case involving Sarbanes-Oxley, for example, an employee possessed a large number of confidential documents, compelling a court to observe that a whistleblowing policy "[b]y no means . . . authorize[s] disgruntled employees to pilfer a wheelbarrow" full of proprietary documents.[520]

Moreover, it bears noting that the risk of such disclosures—even to a trusted intermediary under the DTSA—may lead companies to be ever more protective over their algorithms, limiting exposure to only the most loyal of employees or by overly directing resources towards their constant surveillance and protection.[521] As one lawyer puts it, "[t]he immunity provision of the DTSA

---

514. *Id.* at 38.
515. *Id.* at 9.
516. *See* Barocas & Selbst, *supra* note 60, at 692.
517. *See* Mike Ananny & Kate Crawford, *Seeing Without Knowing: Limitations of the Transparency Ideal and Its Application to Algorithmic Accountability*, 20 NEW MEDIA & SOC'Y 973, 974 (2018).
518. Jenna Burrell, *How the Machine 'Thinks': Understanding Opacity in Machine Learning Algorithms*, 3 BIG DATA & SOC'Y 1–2 (2016).
519. *Id.*
520. JDS Uniphase Corp. v. Jennings, 473 F. Supp. 2d 697, 702 (E.D. Va. 2007). Instead, the court granted a further status conference to consider the extent of the breach, noting that it needed to examine which of the documents were proprietary and the extent of the remedy to be granted. *Id.* at 705.
521. Another objection to a whistleblowing exception stems from FOIA. Although regulatory agencies often compel companies to disclose their trade secrets, either through a contractual protects disclosing individuals, but individuals cannot disclose if they have no access to the information."[522]

Finally, it bears mentioning that this solution is only a partial one—it does not go far enough. For example, immunity under the DTSA would be more effective if coupled with a similar whistleblowing exception to the Computer Fraud and Abuse Act (CFAA) (or even a limited exclusion for terms of service violations).[523] It becomes clear, from the issues surrounding the CFAA, that a whistleblower exception might be further warranted in those circumstances, leading commentators to support the idea.[524] In 2008, in the wake of the suicide of Aaron Swartz, U.S. Representative Zoe Lofgren proposed a bill, called "Aaron's law" that would have excluded terms of service violations from the list of violations under the CFAA.[525] It was never passed. And while the DTSA provides immunity for disclosures to attorneys or government officials, it does not immunize disclosures made to journalists, academics, watchdog groups, or the general public.[526]

Of course, noting the above, it would be an overstatement to say that a whistleblowing exception, as it exists in the DTSA, solves the problem of algorithmic accountability. It cannot solve, as a substantive matter, the issue of how to make algorithms more accountable.[527] However, it would also be an understatement to say that the DTSA's whistleblowing provisions are completely unrelated to the issue of algorithmic transparency. For one thing, they avoid the pitfalls associated with full transparency (like destruction of a trade secret), because they provide for a sort of *in camera* review of the

---

agreement or through regulatory activities, there is some risk that FOIA could be used to circumvent the seclusion that a government could provide. Wilson, *supra* note 448, at 276–77 (discussing this possibility). Although FOIA is designed to encourage disclosure of general information, it includes an exemption for trade secrets. *Id.* at 281. However, case law has suggested that the government has a discretionary ability to disclose trade secrets to a requesting party under certain circumstances. *Id.* Since many federal regulations actually require a government agency to provide notice to the trade secret holder, the trade secret holder can then institute review under the Administrative Procedure Act. If an improper disclosure occurred, the trade secret holder may be able to file a claim for compensation under the Fifth Amendment Takings Clause. *Id.* at 281–82; *see* Ruckelshaus v. Monsanto Co., 467 U.S. 986, 1012(1984).

522. *See* Jordan J. Altman, Doreen E. Lilienfeld & Mark Pereira, *License to Leak: The DTSA and the Risks of Immunity*, INTELL. PROP. & TECH. L.J., Oct. 2016, at 8.

523. *See* Erika Spath, *Whistleblowers Take a Gamble Under the CFAA: How Federal Prosecutors Game the System Under the Proposed Changes to the Act*, 37 U. LA VERNE L. REV. 369, 401–02 (2016).

524. *Id.* at 401 (arguing that Congress should enact a statutory law exception to the CFAA).

525. *Id.* at 373–74.

526. Altman et al. *supra* note 522, at 6–7.

527. *See generally* Kroll et al., *supra* note 60.

information.  Second, because the person bringing the information to the lawyer or official is an employee, he or she may be able to address the substantial issues of information asymmetry associated with external audits, thus addressing issues of technological literacy and complexity.

Indeed, although the DTSA provisions are extraordinarily promising, it is also important to note that other jurisdictions have taken even broader steps to protect whistleblowing.  For example, an equivalent EU Directive included numerous exceptions for the public disclosure of trade secrets "'for exercising the right to freedom of expression and information . . . , including respect for freedom and pluralism of the media,' and for revealing a 'misconduct, wrongdoing or illegal activity, provided that the respondent acted for the purpose of protecting the general public interest.'"[528]

Admittedly, no solution is perfect, due in no small part to the administrative costs involved and the difficulty of detection.  However, by exploiting exemptions in existing law, and by supplementing those exemptions with particular audit requirements, we can create some step towards encouraging a greater culture of algorithmic accountability.  At the very least, the whistleblower exemption should encourage companies to be ever more vigilant about the risks of discrimination, since it demonstrates that secrecy might not always trump accountability.

## Conclusion

This Article has explored both the limits and the possibilities of bringing a culture of accountability to algorithms.  As I have argued, in the absence of oversight, a mixture of industry self-regulation and whistleblower engagement offers us one path forward in the future direction of civil rights law to address the issues raised by AI.  We can no longer afford to consider issues of informational privacy and due process in a vacuum.  Instead of focusing on the value of explanations, we must turn towards lifting the veil of secrecy that allows companies to escape detection.  Or we must incentivize companies to remain vigilant against discrimination through other means.

As this Article suggests, it is indeed possible to exploit the potential for whistleblower liability as a public policy exemption to encourage greater algorithmic transparency.  On a much deeper, more abstract level, the availability of these solutions also portends a much-needed shift in our language and approach to civil rights. The future of civil rights in an age of AI requires us to explore the limitations within intellectual property and, more specifically, trade secrets. If we can exploit the exemptions that already exist within trade secret law, we can also create an entirely new generation of civil rights developments altogether.

---

528.    Anand B. Patel et al., *The Global Harmonization of Trade Secret Law: The Convergence of Protections for Trade Secret Information in the United States and European Union*, 83 Def. Couns. J. 472, 484 (2016) (footnotes omitted).